\documentclass[10pt]{iopart}
\usepackage{epsfig,latexsym}
\usepackage[perpage,symbol*]{footmisc}
\eqnobysec

\def\gtrsim{\;\rlap{\lower 2.5pt
 \hbox{$\sim$}}\raise 1.5pt\hbox{$>$}\;}
\def\lesssim{\;\rlap{\lower 2.5pt
   \hbox{$\sim$}}\raise 1.5pt\hbox{$<$}\;}

\begin{document}
\title[The basics of gravitational wave theory]
{The basics of gravitational wave theory}

\author{\'Eanna \'E Flanagan\dag\ and Scott A Hughes\ddag}
\address{\dag\ Center for Radiophysics and Space Research, Cornell
University, Ithaca, NY 14853}
\address{\ddag\ Department of Physics and Center for Space Research,
Massachusetts Institute of Technology, Cambridge, MA 02139}

\begin{abstract}
Einstein's special theory of relativity revolutionized physics by
teaching us that space and time are not separate entities, but join as
``spacetime''.  His general theory of relativity further taught us
that spacetime is not just a stage on which dynamics takes place, but
is a participant: The field equation of general relativity connects
matter dynamics to the curvature of spacetime.  Curvature is
responsible for gravity, carrying us beyond the Newtonian conception
of gravity that had been in place for the previous two and a half
centuries.  Much research in gravitation since then has explored and
clarified the consequences of this revolution; the notion of dynamical
spacetime is now firmly established in the toolkit of modern physics.
Indeed, this notion is so well established that we may now contemplate
using spacetime as a {\it tool} for other science.  One aspect of
dynamical spacetime --- its radiative character, ``gravitational
radiation'' --- will inaugurate entirely new techniques for observing
violent astrophysical processes.  Over the next one hundred years,
much of this subject's excitement will come from learning how to {\it
exploit} spacetime as a tool for astronomy.  This article is intended
as a tutorial in the basics of gravitational radiation physics.
\end{abstract}

\section{Introduction: Spacetime and gravitational waves}
\label{sec:intro}

Einstein's special relativity {\cite{ae1905}} taught us that space and
time are not simply abstract, external concepts, but must in fact be
considered measured observables, like any other quantity in physics.  This
reformulation {\it enforced} the philosophy that Newton sought to
introduce in laying out his laws of mechanics {\cite{in1687a}:
\begin{quote}
$\ldots$ I frame no hypotheses; for whatever is not reduced from the
phenomena is to be called an hypothesis; and hypotheses $\ldots$ have
no place in experimental philosophy $\ldots$
\end{quote}
Despite his intention to stick only with that which can be observed,
Newton described space and time using exactly the abstract notions
that he otherwise deplored {\cite{in1687b}}:
\begin{quote}
Absolute space, in its own nature, without relation to anything
external, remains always similar and immovable
\end{quote}
\begin{quote}
Absolute, true, and mathematical time, of itself, and from its own
nature, flows equably without relation to anything external.
\end{quote}

Special relativity put an end to these abstractions: Time is nothing
more than that which is measured by clocks, and space is that which is
measured by rulers.  The properties of space and time thus depend on
the properties of clocks and rulers.  The constancy of the speed of
light as measured by observers in different reference frames, as
observed in the Michelson-Morley experiment, forces us inevitably to
the fact that space and time are mixed into spacetime.  Ten years
after his paper on special relativity, Einstein endowed spacetime with
curvature and made it dynamical {\cite{ae1915a}}.  This provided a
{\it covariant} theory of gravity {\cite{ae1915b}, in which all
predictions for physical measurements are invariant under changes in
coordinates.  In this theory, general relativity, the notion of
``gravitational force'' is reinterpreted in terms of the behavior of
geodesics in the curved manifold of spacetime.

To be compatible with special relativity, gravity must be causal: Any
change to a gravitating source must be communicated to distant
observers no faster than the speed of light, $c$.  This leads
immediately to the idea that there must exist some notion of
``gravitational radiation''.  As demonstrated by Bernard Schutz, one
can actually calculate with surprising accuracy many of the properties
of gravitational radiation simply by combining a time dependent
Newtonian potential with special relativity {\cite{bs1984}}.

The first calculation of gravitational radiation in general relativity
is due to Einstein.  His initial calculation {\cite{ae1916}} was
``marred by an error in calculation'' (Einstein's words), and was
corrected in 1918 {\cite{ae1918}} (albeit with an overall factor of
two error).  Modulo a somewhat convoluted history (discussed in great
detail by Kennefick {\cite{dk1997}}) owing (largely) to the
difficulties of analyzing radiation in a nonlinear theory, Einstein's
final result stands today as the leading-order ``quadrupole formula'' for
gravitational wave emission.  This formula plays a role in gravity
theory analogous to the dipole formula for electromagnetic radiation,
showing that gravitational waves (hereafter abbreviated GWs) arise
from accelerated masses exactly as electromagnetic waves arise from
accelerated charges.

The quadrupole formula tells us that GWs are difficult to produce ---
very large masses moving at relativistic speeds are needed.  This
follows from the weakness of the gravitational interaction.  A
consequence of this is that it is {\it extremely} unlikely there will
ever be an interesting laboratory source of GWs.  The only objects
massive and relativistic enough to generate detectable GWs are
astrophysical.  Indeed, experimental confirmation of the existence of
GWs has come from the study of binary neutron star systems --- the
variation of the mass quadrupole in such systems is large enough that
GW emission changes the system's characteristics on a timescale short
enough to be observed.  The most celebrated example is the
``Hulse-Taylor'' pulsar, B1913+16, reported by Russell Hulse and Joe
Taylor in 1975 {\cite{ht75}}.  Thirty years of observation have shown
that the orbit is decaying; the results match with extraordinary
precision general relativity's prediction for such a decay due to the
loss of orbital energy and angular momentum by GWs.  For a summary of
the most recent data, see Ref.\ {\cite{wt04}}, Fig.\ 1.  Hulse and
Taylor were awarded the Nobel Prize for this discovery in 1993.  Since
this pioneering system was discovered, several other double neutron
star systems ``tight'' enough to exhibit strong GW emission have been
discovered {\cite{sttw02,dk94,betal03,fetal04}}.

Studies of these systems prove beyond a reasonable doubt that GWs
exist.  What remains is to detect the waves directly and exploit them
--- to {\it use} GWs as a way to study astrophysical objects.  The
contribution to this volume by Danzmann {\cite{kd_here}} discusses the
challenges and program of directly measuring these waves.
Intuitively, it is clear that measuring these waves must be difficult
--- the weakness of the gravitational interaction ensures that the
response of any detector to gravitational waves is very small.
Nonetheless, technology has brought us to the point where detectors
are now beginning to set interesting upper limits on GWs from some
sources {\cite{ligo1,ligo2,ligo3,ligo4}}.  First direct detection is
now hopefully not too far in the future.

The real excitement will come when detection becomes routine.  We will
then have an opportunity to retool the ``physics experiment'' of
direct detection into the development of astronomical observatories.
Some of the articles appearing in this volume will discuss likely
future revolutions which, at least conceptually, should change our
notions of spacetime in a manner as fundamental as Einstein's works in
1905 and 1915 (see, e.g., papers by Ashtekar and Horowitz).  Such a
revolution is unlikely to be driven by GW observations --- most of
what we expect to learn using GWs will apply to regions of spacetime
that are well-described using classical general relativity; it is
highly unlikely that Einstein's theory will need major revisions
prompted by GW observations.  Any revolution arising from GW science
will instead be in astrophysics: Mature GW measurements have the
potential to study regions of the Universe that are currently
inaccessible to our instruments.  During the next century of spacetime
study, spacetime will itself be exploited to study our Universe.

\subsection{Why this article?}

As GW detectors have improved and approached maturity, many articles
have been written reviewing this field and its promise.  One might
ask: Do we really need another one?  As partial answer to this
question, we note that Richard Price requested this article very
nicely.  More seriously, our goal is to provide a brief tutorial on
the basics of GW science, rather than a comprehensive survey of the
field.  The reader who is interested in such a survey can find such in
Refs.\
{\cite{bm94,bs99,glpps00,snowmass,bs02,ct02,finn-99,annals,lg03}}.
Other reviews on the basics of GW science can be found in Refs.\
{\cite{thorne1983,schutz2001}}; we also recommend the dedicated
conference procedings {\cite{leshouches,ciufolini,amaldi5}}.

We assume that the reader has a basic familiarity with general
relativity, at least at the level of Hartle's textbook {\cite{jh03}};
thus, we assume the reader understands metrics and is reasonably
comfortable taking covariant derivatives.  We adapt what Baumgarte and
Shapiro {\cite{bs03}} call the ``Fortran convention'' for indices:
$a,b,c \ldots h$ denote spacetime indices which run over $0,1,2,3$ or
$t,x,y,z$, while $i,j,k \ldots n$ denote spatial indices which run
over $1,2,3$.  We use the Einstein summation convention throughout ---
repeated adjacent indices in the ``upstairs'' and ``downstairs''
positions imply a sum:
\[
u^a v_a \equiv \sum_a u^a v_a\;.
\]
When we discuss linearized theory, we will sometimes be sloppy and sum
over adjacent spatial indices in the same position.  Hence,
\[
u_i v_i \equiv u^i v^i \equiv u^i v_i \equiv \sum_i u^i v_i
\]
is valid in linearized theory.  (As we will discuss in Sec.\
{\ref{sec:basic_lin}}, this is allowable because, in linearized
theory, the position of a spatial index is immaterial in Cartesian
coordinates.)  A quantity that is symmetrized on pairs of indices is
written
\[
A_{(ab)} = \frac{1}{2}\left(A_{ab} + A_{ba}\right)\;.
\]
Throughout most of this article we use ``relativist's units'', in
which $G = 1 = c$; mass, space, and time have the same units in this
system.  The following conversion factors are often useful for
converting to ``normal'' units:
\begin{eqnarray*}
1\,\mbox{second} &=& 299,792,458\,\mbox{meters} \simeq
3\times10^8\,\mbox{meters}
\\
1\,M_\odot &=& 1476.63\,\mbox{meters} \simeq 1.5\,\mbox{kilometers}
\\
&=& 4.92549\times10^{-6}\,\mbox{seconds} \simeq
5\,\mbox{microseconds}\;.
\end{eqnarray*}
($1\,M_\odot$ is one solar mass.)  We occasionally restore factors of
$G$ and $c$ to write certain formulae in normal units.

Section {\ref{sec:basic_lin}} provides an introduction to linearized
gravity, deriving the most basic properties of GWs.  Our treatment in
this section is mostly standard.  One aspect of our treatment that is
slightly unusual is that we introduce a gauge-invariant formalism that
fully characterizes the linearized gravity's degrees of freedom.  We
demonstrate that the linearized Einstein equations can be written as 5
Poisson-type equations for certain combinations of the spacetime
metric, plus a wave equation for the transverse-traceless components
of the metric perturbation.  This analysis helps to clarify which
degrees of freedom in general relativity are radiative and which are
not, a useful exercise for understanding spacetime dynamics.

Section \ref{sec:detect} analyses the interaction of GWs with
detectors whose sizes are small compared to the wavelength of the GWs.
This includes ground-based interferometric and resonant-mass
detectors, but excludes space-based interferometric detectors.  The
analysis is carried out in two different gauges; identical results are
obtained from both analyses.  Section {\ref{sec:lin_with_source}}
derives the leading-order formula for radiation from slowly-moving,
weakly self-gravitating sources, the quadrupole formula discussed
above.

In Sec.\ {\ref{sec:lin_in_curved}}, we develop linearized theory on a
curved background spacetime.  Many of the results of ``basic
linearized theory'' (Sec.\ {\ref{sec:basic_lin}}) carry over with
slight modification.  We introduce the ``geometric optics'' limit in
this section, and sketch the derivation of the Isaacson stress-energy
tensor, demonstrating how GWs carry energy and curve spacetime.
Section {\ref{sec:synopsis}} provides a very brief synopsis of GW
astronomy, leading the reader through a quick tour of the relevant
frequency bands and anticipated sources.  We conclude by discussing
very briefly some topics that we could not cover in this article, with
pointers to good reviews.

\section{The basic basics: Gravitational waves in linearized gravity}
{\label{sec:basic_lin}

The most natural starting point for any discussion of GWs is {\it
linearized gravity}.  Linearized gravity is an adequate approximation
to general relativity when the spacetime metric, $g_{ab}$, may be
treated as deviating only slightly from a flat metric, $\eta_{ab}$:
\begin{equation}
g_{ab} = \eta_{ab} + h_{ab},\qquad ||h_{ab}|| \ll 1\;.
\end{equation}
Here $\eta_{ab}$ is defined to be $\mbox{diag}(-1,1,1,1)$ and
$||h_{ab}||$ means ``the magnitude of a typical non-zero component of
$h_{ab}$''.  Note that the condition $||h_{ab}|| \ll 1$ requires both
the gravitational field to be weak, and in addition constrains the
coordinate system to be approximately Cartesian.  We will refer to
$h_{ab}$ as the metric perturbation; as we will see, it encapsulates
GWs, but contains additional, non-radiative degrees of freedom as
well.  In linearized gravity, the smallness of the perturbation means
that we only keep terms which are linear in $h_{ab}$ --- higher order
terms are discarded.  As a consequence, indices are raised and lowered
using the flat metric $\eta_{ab}$.  The metric perturbation $h_{ab}$
transforms as a tensor under Lorentz transformations, but not under
general coordinate transformations.

We now compute all the quantities which are needed to describe
linearized gravity.  The components of the affine connection
(Christoffel coefficients) are given by
\begin{eqnarray}
{\Gamma^a}_{bc} &=& \frac{1}{2}\eta^{ad}\left(\partial_c h_{db} +
\partial_b h_{dc} -\partial_d h_{bc}\right)
\nonumber\\
&=& \frac{1}{2}\left(\partial_c {h^a}_b + \partial_b {h^a}_{c}
-\partial^a h_{bc}\right)\;.
\label{eq:connection}
\end{eqnarray}
Here $\partial_a$ means the partial derivative $\partial / \partial x^a$.
Since we use $\eta_{ab}$ to raise and lower indices, spatial indices
can be written either in the ``up'' position or the ``down'' position
without changing the value of a quantity: $f^x = f_x$.  Raising or
lowering a time index, by contrast, switches sign: $f^t = -f_t$.  The
Riemann tensor we construct in linearized theory is then given by
\begin{eqnarray}
R^a_{\ bcd} &=& \partial_c{\Gamma^a}_{bd} - \partial_d{\Gamma^a}_{bc}
\nonumber\\
&=& \frac{1}{2}\left(\partial_c\partial_b {h^a}_d +
\partial_d\partial^a h_{bc} - \partial_c\partial^a h_{bd} -
\partial_d\partial_b {h^a}_c\right)\;.
\label{eq:riemann}
\end{eqnarray}
From this, we construct the Ricci tensor
\begin{equation}
R_{ab} = {R^c}_{acb}
= \frac{1}{2}\left(\partial_c\partial_b {h^c}_a + \partial^c
\partial_a h_{bc} - \Box h_{ab} - \partial_a\partial_b h\right)\;,
\label{eq:ricci}
\end{equation}
where $h = {h^a}_a$ is the trace of the metric perturbation, and $\Box
= \partial_c\partial^c = \nabla^2 - \partial_t^2$ is the wave
operator.  Contracting once more, we find the curvature scalar:
\begin{equation}
R = {R^a}_a = \left(\partial_c\partial^a {h^c}_a - \Box h\right)
\label{eq:scalar}
\end{equation}
and finally build the Einstein tensor:
\begin{eqnarray}
G_{ab} &=& R_{ab} - \frac{1}{2}\eta_{ab} R
\nonumber\\
&=& \frac{1}{2}\left(\partial_c\partial_b {h^c}_a + \partial^c
\partial_a h_{bc} - \Box h_{ab} - \partial_a\partial_b h
\right.\nonumber\\
& &\qquad\left.-\eta_{ab}\partial_c\partial^d {h^c}_d + \eta_{ab} \Box
h\right)\;.
\label{eq:einstein_h}
\end{eqnarray}

This expression is a bit unwieldy.  Somewhat remarkably, it can be
cleaned up significantly by changing notation: Rather than working
with the metric perturbation $h_{ab}$, we use the {\it trace-reversed}
perturbation $\bar h_{ab} = h_{ab} - \frac{1}{2}\eta_{ab} h$.  (Notice
that $\bar {h^a}_a = -h$, hence the name ``trace reversed''.)
Replacing $h_{ab}$ with $\bar h_{ab} + \frac{1}{2}\eta_{ab} h$ in Eq.\
(\ref{eq:einstein_h}) and expanding, we find that all terms with the
trace $h$ are canceled.  What remains is
\begin{eqnarray}
G_{ab} = \frac{1}{2}\left(\partial_c\partial_b {{\bar h}^c}_{\ a} +
\partial^c \partial_a \bar h_{bc} - \Box \bar h_{ab} -\eta_{ab}
\partial_c\partial^d {{\bar h}^c}_{\ d}\right)\;.
\label{eq:einstein_hbar}
\end{eqnarray}

This expression can be simplified further by choosing an appropriate
coordinate system, or {\it gauge}.  Gauge transformations in general
relativity are just coordinate transformations.  A general
infinitesimal coordinate transformation can be written as ${x^a}' =
x^a + \xi^a$, where $\xi^a(x^b)$ is an arbitrary infinitesimal vector
field.  This transformation changes the metric via
\begin{equation}
h_{ab}' = h_{ab} - 2\partial_{(a} \xi_{b)}\;,
\label{eq:metric_transform}
\end{equation}
so that the trace-reversed metric becomes
\begin{eqnarray}
\bar h_{ab}' &=& h_{ab}' - \frac{1}{2}\eta_{ab} h'
\nonumber\\
&=& \bar h_{ab} - 2\partial_{(b}\xi_{a)} + \eta_{ab}\partial^c\xi_c\;.
\label{eq:barmetric_transform}
\end{eqnarray}
A class of gauges that are commonly used in studies of radiation are
those satisfying the {\it Lorentz gauge} condition
\begin{equation}
\partial^a \bar h_{ab} = 0\;.
\label{eq:lorentzgauge}
\end{equation}
(Note the close analogy to Lorentz gauge\footnote{Fairly recently, it
has become widely recognized that this gauge was in fact invented by
Ludwig Lorenz, rather than by Hendrik Lorentz.  The inclusion of the
``t'' seems most likely due to confusion between the similar names;
see Ref.\ {\cite{vanbladel}} for detailed discussion.  Following the
practice of Griffiths ({\cite{griffiths}}, p.\ 421), we bow to the
weight of historical usage in order to avoid any possible confusion.}
in electromagnetic theory, $\partial^a A_a = 0$, where $A_a$ is the
potential vector.)

Suppose that our metric perturbation is not in Lorentz gauge.  What
properties must $\xi_a$ satisfy in order to {\it impose} Lorentz
gauge?  Our goal is to find a new metric $h'_{ab}$ such that
$\partial^a \bar h'_{ab} = 0$:
\begin{eqnarray}
\partial^a \bar h_{ab}' &=& \partial^a \bar h_{ab} -
\partial^a\partial_b\xi_a - \Box \xi_b + \partial_b \partial^c\xi_c
\\
&=& \partial^a \bar h_{ab} - \Box\xi_b\;.
\label{eq:ll1}
\end{eqnarray}
Any metric perturbation $h_{ab}$ can therefore be put into a Lorentz
gauge by making an infinitesimal coordinate transformation that satisfies
\begin{equation}
\Box \xi_b = \partial^a \bar h_{ab}\;.
\label{eq:wave}
\end{equation}
One can always find solutions to the wave equation (\ref{eq:wave}),
thus achieving Lorentz gauge.
The amount of gauge freedom has now been reduced from 4 freely
specifiable functions of 4 variables to 4 functions of 4 variables
that satisfy the homogeneous wave equation $\Box \xi^b =0$, or,
equivalently, to 8 freely specifiable functions of 3 variables on an
initial data hypersurface.

Applying the Lorentz gauge condition (\ref{eq:lorentzgauge}) to the
expression (\ref{eq:einstein_hbar}) for the Einstein tensor, we find
that all but one term vanishes:
\begin{eqnarray}
G_{ab} = -\frac{1}{2}\Box \bar h_{ab}\;.
\label{eq:einstein_lg}
\end{eqnarray}
Thus, in Lorentz gauges, the Einstein tensor simply reduces to the
wave operator acting on the trace reversed metric perturbation (up to
a factor $-1/2$).  The linearized Einstein equation is therefore
\begin{equation}
\Box \bar h_{ab} = - 16 \pi T_{ab}\;;
\label{eq:elin}
\end{equation}
in vacuum, this reduces to
\begin{equation}
\Box \bar h_{ab} = 0\;.
\label{eq:elin1}
\end{equation}
Just as in electromagnetism, the equation (\ref{eq:elin}) admits a
class of homogeneous solutions which are superpositions of plane
waves:
\begin{equation}
{\bar h}_{ab}({\bf x},t) = {\rm Re} \int d^3 k \ A_{ab}({\bf k}) e^{i
  ({\bf k} \cdot {\bf x} - \omega t)}\;.
\label{eq:planewaves}
\end{equation}
Here, $\omega = |{\bf k}|$.  The complex coefficients $A_{ab}({\bf
k})$ depend on the wavevector ${\bf k}$ but are independent of ${\bf
x}$ and $t$.  They are subject to the constraint $k^a A_{ab} = 0$
(which follows from the Lorentz gauge condition), with $k^a =
(\omega,{\bf k})$, but are otherwise arbitrary.  These solutions are
gravitational waves.

\subsection{Globally vacuum spacetimes: Transverse traceless (TT) gauge}
\label{sec:TTgauge}

We now specialize to globally vacuum spacetimes in which $T_{ab}=0$
everywhere, and which are asymptotically flat (for our purposes,
$h_{ab} \to 0$ as $r \to \infty$).  Equivalently, we specialize to the
space of homogeneous, asymptotically flat solutions of the linearized
Einstein equation (\ref{eq:elin}).  For such spacetimes one can, along
with choosing Lorentz gauge, further specialize the gauge to make the
metric perturbation be purely spatial
\begin{equation}
h_{tt} = h_{ti} =0
\label{eq:spatial}
\end{equation}
and traceless
\begin{equation}
h = h_i^{\ i} =0.
\label{eq:traceless}
\end{equation}
The Lorentz gauge condition (\ref{eq:lorentzgauge}) then
implies that the spatial metric perturbation is transverse:
\begin{equation}
\partial_i h_{ij} =0.
\label{eq:transverse}
\end{equation}
This is called the transverse-traceless gauge, or TT gauge.  A metric
perturbation that has been put into TT gauge will be written
$h_{ab}^{\rm TT}$.  Since it is traceless, there is no distinction
between $h^{\rm TT}_{ab}$ and $\bar h^{\rm TT}_{ab}$.

The conditions (\ref{eq:spatial}) and (\ref{eq:traceless}) comprise 5
constraints on the metric, while the residual gauge freedom in Lorentz
gauge is parameterized by 4 functions that satisfy the wave equation.
It is nevertheless possible to satisfy these conditions,
essentially because the metric perturbation satisfies the linearized
vacuum Einstein equation.  When the TT gauge conditions are satisfied
the gauge is completely fixed.

One might wonder {\it why} we would choose TT gauge.  It is certainly
not necessary; however, it is extremely {\it convenient}, since the TT
gauge conditions completely fix all the local gauge freedom.  The
metric perturbation $h_{ab}^{\rm TT}$ therefore contains only
physical, non-gauge information about the radiation.  In TT gauge
there is a close relation between the metric perturbation and the
linearized Riemann tensor $R_{abcd}$ [which is invariant under the
local gauge transformations (\ref{eq:metric_transform}) by Eq.\
(\ref{eq:riemann})], namely
\begin{equation}
R_{itjt} = - \frac{1}{2} {\ddot h}_{ij}^{\rm TT}.
\label{eq:Rtitj}
\end{equation}
In a globally vacuum spacetime, all non-zero components of the Riemann
tensor can be obtained from $R_{itjt}$ via Riemann's symmetries
and the Bianchi identity.  In a
more general spacetime, there will be components that are not related
to radiation; this point is discussed further in Sec.\
{\ref{subsec:gauge_invar}}.

Transverse traceless gauge also exhibits the fact that gravitational
waves have two polarization components.  For example, consider a GW
which propagates in the $z$ direction: $h^{\rm TT}_{ij} = h^{\rm
TT}_{ij}(t - z)$ is a valid solution to the wave equation $\Box h^{\rm
TT}_{ij} = 0$.  The Lorentz condition $\partial_z h^{\rm TT}_{zj} = 0$
implies that $h^{\rm TT}_{zj}(t - z) = \mbox{constant}$.  This
constant must be zero to satisfy the condition that $h_{ab} \to 0$ as
$r \to \infty$.  The only non-zero components of $h^{\rm TT}_{ij}$ are
then $h^{\rm TT}_{xx}$, $h^{\rm TT}_{xy}$, $h^{\rm TT}_{yx}$, and
$h^{\rm TT}_{yy}$.  Symmetry and the tracefree condition
(\ref{eq:traceless}) further mandate that only two of these are
independent:
\begin{eqnarray}
h^{\rm TT}_{xx} &=& -h^{\rm TT}_{yy} \equiv h_+(t-z)\;;
\\
h^{\rm TT}_{xy} &=& h^{\rm TT}_{yx} \equiv h_\times(t-z)\;.
\label{eq:pols0}
\end{eqnarray}
The quantities $h_+$ and $h_\times$ are the two independent waveforms
of the GW (see Fig.\ \ref{fig:forcelines}).

For globally vacuum spacetimes, one can always satisfy the TT gauge
conditions.  To see this, note that the most general gauge
transformation $\xi^a$ that preserves the Lorentz gauge condition
(\ref{eq:lorentzgauge}) satisfies $\Box \xi^a =0$, from Eq.\
(\ref{eq:ll1}).  The general solution to this equation can be written
\begin{equation}
\xi^a = {\rm Re} \int d^3 k \ C^a({\bf k}) e^{i ({\bf k} \cdot {\bf x}
- \omega t)}
\end{equation}
for some coefficients $C^a({\bf k})$.  Under this transformation the
tensor $A_{ab}({\bf k})$ in Eq.\ (\ref{eq:planewaves}) transforms as
\begin{equation}
A_{ab} \to A'_{ab}=A_{ab} - 2 i k_{(a} C_{b)} + i \eta_{ab} k^d C_d\;.
\end{equation}
Achieving the TT gauge conditions (\ref{eq:transverse}) and
(\ref{eq:traceless}) therefore requires finding, for each ${\bf k}$, a
$C^a({\bf k})$ that satisfies the two equations
\begin{eqnarray}
0 &=& \eta^{ab} A'_{ab} = \eta^{ab} A_{ab} + 2 i k^a C_a \\ \mbox{} 0
&=& A'_{tb} = A_{tb} - i C_b k_t - i C_t k_b + i {\delta^t}_b (k^a
C_a)\;;
\end{eqnarray}
${\delta^t}_b$ is the Kronecker delta --- zero for $b \ne t$, unity
otherwise.  An explicit solution to these equations is given by
\begin{equation}
C_a = \frac{3 A_{bc} l^b l^c}{8 i \omega^4} k_a + \frac{\eta^{bc}
  A_{bc} }{4 i \omega^4} l_a + \frac{1}{2 i \omega^2} A_{ab} l^b\;,
\end{equation}
where $k^a = (\omega,{\bf k})$ and $l^a = (\omega,-{\bf k})$.

\subsection{Global spacetimes with matter sources}
\label{subsec:gauge_invar}

We now return to the more general and realistic situation in which the
stress-energy tensor is non-zero.  We continue to assume that the
linearized Einstein equations are valid everywhere in spacetime, and
that we consider asymptotically flat solutions only.  In this context,
the metric perturbation $h_{ab}$ contains (i) gauge degrees of
freedom; (ii) physical, radiative degrees of freedom; and (iii)
physical, non-radiative degrees of freedom tied to the matter sources.
Because of the presence of the physical, non-radiative degrees of
freedom, it is not possible in general to write the metric
perturbation in TT gauge.  However, the metric perturbation can be
split up uniquely into various pieces that correspond to the degrees
of freedom (i), (ii) and (iii), and the radiative degrees of freedom
correspond to a piece of the metric perturbation that satisfies the TT
gauge conditions, the so-called TT piece.

This aspect of linearized theory is obscured by the standard, Lorentz
gauge formulation (\ref{eq:elin}) of the linearized Einstein
equations.  There, all the components of $h_{ab}$ appear to be
radiative, since all the components obey wave equations.  In this
subsection, we describe a formulation of linearized theory which
focuses on gauge invariant observables.  In particular, we will see
that {\it only} the TT part of the metric obeys a wave equation in all
gauges.  We show that the non-TT parts of the metric can be gathered
into a set of gauge invariant functions; these functions are governed
by Poisson equations rather than wave equations.  This shows that {\it
the non-TT pieces of the metric do not exhibit radiative degrees of
freedom.}  Although one can always choose gauges like Lorentz gauge in
which the non-radiative parts of the metric obey wave equations and
thus {\it appear} to be radiative, this appearance is gauge artifact.
Such gauge choices, although useful for calculations, can cause one to
mistake purely gauge modes for truly physical radiation.

Interestingly, the first analysis contrasting physical radiative
degrees of freedom from purely coordinate modes appears to have been
performed by Eddington in 1922 {\cite{ae1922}}.  Eddington was
somewhat suspicious of Einstein's analysis {\cite{ae1918}}, as
Einstein chose a gauge in which all metric functions propagated with
the speed of light.  Though the entire metric appeared to be radiative
(by construction), Einstein found that {\it only} the
``transverse-transverse'' pieces of the metric carried energy.
Eddington wrote:
\begin{quote}
Weyl has classified plane gravitational waves into three types,
viz.~(1) longitudinal-longitudinal; (2) longitudinal-transverse; (3)
transverse-trans-verse.  The present investigation leads to the
conclusion that transverse-transverse waves are propagated with the
speed of light {\it in all systems of co-ordinates}.  Waves of the
first and second types have no fixed velocity --- a result which
rouses suspicion as to their objective existence.  Einstein had also
become suspicious of these waves (in so far as they occur in his
special co-ordinate system) for another reason, because he found that
they convey no energy.  They are not objective, and (like absolute
velocity) are not detectable by any conceivable experiment.  They are
merely sinuosities in the co-ordinate system, and the only speed of
propagation relevant to them is ``the speed of thought.''

$\ldots$ It is evidently a great convenience in analysis to have all
waves, both physical and spurious, travelling with one velocity; but
it is liable to obscure physical ideas by mixing them up so
completely.  The chief new point in the present discussion is that
when unrestricted co-ordinates are allowed the genuine waves continue
to travel with the velocity of light and the spurious waves cease to
have any fixed velocity.
\end{quote}
Unfortunately, Eddington's wry dismissal of unphysical modes as
propagating with ``the speed of thought'' is often taken by skeptics
(and crackpots) as applying to {\it all} gravitational perturbations.
Eddington in fact showed quite the opposite.  We do so now using
somewhat more modern notation; our presentation is essentially the
flat-spacetime limit of Bardeen's {\cite{jb1980}} gauge-invariant
cosmological perturbation formalism.  A similar treatment can be
found in lecture notes by Bertschinger {\cite{eb1999}}.

We begin by defining the decomposition of the metric perturbation
$h_{ab}$, in any gauge, into a number of irreducible pieces.
Assuming that $h_{ab} \to 0$ as $r \to \infty$, we define the
quantities $\phi$, $\beta_i$, $\gamma$, $H$, $\varepsilon_i$,
$\lambda$ and $h_{ij}^{\rm TT}$ via the equations
\begin{eqnarray}
h_{tt} &=& 2\phi\;,
\label{eq:htt_invar}\\
h_{ti} &=& \beta_i + \partial_i\gamma\;,
\label{eq:hti_invar}\\
h_{ij} &=& h_{ij}^{\rm TT} + \frac{1}{3}H\delta_{ij} +
\partial_{(i}\varepsilon_{j)} + \left(\partial_i\partial_j -
\frac{1}{3}\delta_{ij}\nabla^2\right)\lambda\;,
\label{eq:hij_invar}
\end{eqnarray}
together with the constraints
\begin{eqnarray}
\partial_i\beta_i &=& 0\quad\mbox{(1 constraint)}
\label{eq:constraint1}\\
\partial_i\varepsilon_i &=& 0\quad\mbox{(1 constraint)}
\label{eq:constraint2}\\
\partial_i h_{ij}^{\rm TT} &=& 0\quad\mbox{(3 constraints)}
\label{eq:constraint3}\\
\delta^{ij}h_{ij}^{\rm TT} &=& 0\quad\mbox{(1 constraint)}
\label{eq:constraint4}
\end{eqnarray}
and boundary conditions
\begin{equation}
\gamma \to 0,\ \ \  \varepsilon_i \to 0,\ \ \  \lambda \to 0,\ \ \
\nabla^2 \lambda \to 0
\label{eq:boundary}
\end{equation}
as $r \to \infty$.  Here $H \equiv \delta^{ij} h_{ij}$ is the trace of
the {\it spatial} portion of the metric perturbation, not to be
confused with the spacetime trace $h = \eta^{ab}h_{ab}$ that we used
earlier.  The spatial tensor $h_{ij}^{\rm TT}$ is transverse and
traceless, and is the TT piece of the metric discussed above which
contains the physical radiative degrees of freedom.  The quantities
$\beta_i$ and $\partial_i \gamma$ are the transverse and longitudinal
pieces of $h_{ti}$.  The uniqueness of this decomposition follows from
taking a divergence of Eq.\ (\ref{eq:hti_invar}) giving $\nabla^2
\gamma = \partial_i h_{ti}$, which has a unique solution by the
boundary condition (\ref{eq:boundary}).  Similarly, taking two
derivatives of Eq.\ (\ref{eq:hij_invar}) yields the equation $2
\nabla^2 \nabla^2 \lambda = 3 \partial_i \partial_j h_{ij} - \nabla^2
H$, which has a unique solution by Eq.\ (\ref{eq:boundary}).  Having
solved for $\lambda$, one can obtain a unique $\varepsilon_i$ by
solving $3 \nabla^2 \varepsilon_i = 6 \partial_j h_{ij} - 2 \partial_i
H - 4 \partial_i \nabla^2 \lambda$.

The total number of free functions in the parameterization
(\ref{eq:htt_invar}) -- (\ref{eq:hij_invar})
of the metric is 16: 4 scalars ($\phi$, $\gamma$, $H$, and $\lambda$), 6
vector components ($\beta_i$ and $\varepsilon_i$), and 6 symmetric
tensor components ($h_{ij}^{\rm TT}$).  The number of constraints
(\ref{eq:constraint1}) -- (\ref{eq:constraint4}) is 6, so the number
of independent variables in the parameterization is 10, consistent
with a symmetric $4\times4$
tensor.

We next discuss how the variables
$\phi$, $\beta_i$, $\gamma$, $H$, $\varepsilon_i$,
$\lambda$ and $h_{ij}^{\rm TT}$
transform under gauge transformations $\xi^a$ with $\xi^a \to 0$ as $r
\to \infty$.
We parameterize such gauge transformation as
\begin{equation}
\xi_a = (\xi_t,\xi_i) \equiv (A,B_i + \partial_i C)\;,
\label{eq:gauge_functions}
\end{equation}
where $\partial_i B_i = 0$ and $C \to 0$ as $r \to \infty$;
thus $B_i$ and $\partial_i C$ are the transverse and longitudinal
pieces of the spatial gauge transformation.
The transformed metric is $h_{ab} -
2\partial_{(a}\xi_{b)}$; decomposing this transformed metric into its
irreducible pieces yields the transformation laws
\begin{eqnarray}
\phi &\to& \quad \phi - \dot A\;,
\label{eq:phi_gauge}\\
\beta_i &\to& \quad \beta_i - \dot B_i\;,
\label{eq:beta_gauge}\\
\gamma &\to& \quad \gamma - A - \dot C\;,
\label{eq:gamma_gauge}\\
H &\to& \quad H - 2\nabla^2C\;,
\label{eq:h_gauge}\\
\lambda &\to& \quad \lambda - 2 C\;,
\label{eq:lambda_gauge}\\
\varepsilon_i &\to& \quad \varepsilon_i - 2 B_i\;,
\label{eq:epsilon_gauge}\\
h_{ij}^{\rm TT} &\to& \quad h_{ij}^{\rm TT}\;.
\label{eq:hTT_gauge}
\end{eqnarray}
Gathering terms, we see that the following combinations of these
functions are gauge invariant:
\begin{eqnarray}
\Phi &\equiv& -\phi + \dot\gamma - \frac{1}{2}\ddot\lambda\;,
\label{eq:Phidef}\\
\Theta &\equiv& \frac{1}{3}\left(H - \nabla^2\lambda\right)\;,
\label{eq:Thetadef}\\
\Xi_i &\equiv& \beta_i - \frac{1}{2}\dot\varepsilon_i\;;
\label{eq:Xidef}
\end{eqnarray}
$h_{ij}^{\rm TT}$ is gauge-invariant without any further manipulation.
In the Newtonian limit $\Phi$ reduces to the Newtonian potential
$\Phi_N$, while $\Theta = - 2 \Phi_N$.  The total number of free,
gauge-invariant functions is 6: 1 function $\Theta$; 1 function
$\Phi$; 3 functions $\Xi_i$, minus 1 due to the constraint
$\partial_i\Xi_i = 0$; and 6 functions $h_{ij}^{\rm TT}$, minus 3 due
to the constraints $\partial_i h_{ij}^{\rm TT} = 0$, minus 1 due to
the constraint $\delta^{ij} h_{ij}^{\rm TT} = 0$.  This is in keeping
with the fact that in general the 10 metric functions contain 6
physical and 4 gauge degrees of freedom.

We would now like to enforce Einstein's equation.  Before doing so, it
is useful to first decompose the stress energy tensor in a manner
similar to that of our decomposition of the metric.  We define the
quantities $\rho$, $S_i$, $S$, $P$, $\sigma_{ij}$, $\sigma_i$ and
$\sigma$ via the equations
\begin{eqnarray}
T_{tt} &=& \rho\;,
\\
T_{ti} &=& S_i + \partial_i S\;,
\\
T_{ij} &=& P\delta_{ij} + \sigma_{ij} + \partial_{(i}\sigma_{j)} +
\left(\partial_i\partial_j -
\frac{1}{3}\delta_{ij}\nabla^2\right)\sigma,\;
\end{eqnarray}
together with the constraints
\begin{eqnarray}
\partial_i S_i &=& 0\;,
\\
\partial_i\sigma_i &=& 0\;,
\\
\partial_i\sigma_{ij} &=& 0\;,
\\
\delta^{ij}\sigma_{ij} &=& 0,\;
\end{eqnarray}
and boundary conditions
\begin{equation}
S \to 0,\ \ \  \sigma_i \to 0,\ \ \  \sigma \to 0,\ \ \
\nabla^2 \sigma \to 0
\label{eq:boundary1}
\end{equation}
as $r \to \infty$.  These quantities are not all independent.  The
variables $\rho$, $P$, $S_i$ and $\sigma_{ij}$ can be specified
arbitrarily; stress-energy conservation ($\partial^a T_{ab} = 0$) then
determines the remaining variables $S$, $\sigma$, and $\sigma_i$ via
\begin{eqnarray}
\nabla^2 S &=& \dot\rho\;,
\label{eq:Sconstraint}\\
\nabla^2 \sigma &=& -\frac{3}{2}P + \frac{3}{2}\dot S\;,
\label{eq:sigma_constraint}\\
\nabla^2\sigma_i &=& 2\dot S_i\;.
\label{eq:sigma_i_constraint}
\end{eqnarray}

We now compute the Einstein tensor from the metric
(\ref{eq:htt_invar}) -- (\ref{eq:hij_invar}).  The result can be
expressed in terms of the gauge invariant observables:
\begin{eqnarray}
G_{tt} &=& -\nabla^2\Theta\;,
\label{eq:Gtt}\\
G_{ti} &=& -\frac{1}{2}\nabla^2\Xi_i - \partial_i\dot\Theta\;,
\label{eq:Gti}\\
G_{ij} &=& -\frac{1}{2}\Box h_{ij}^{\rm TT} -
\partial_{(i}\dot\Xi_{j)} - \frac{1}{2}\partial_i\partial_j\left(2\Phi
+ \Theta\right)\nonumber\\
& &\qquad + \delta_{ij}\left[\frac{1}{2}\nabla^2
\left(2\Phi + \Theta\right) - \ddot \Theta\right]\;.
\label{eq:Gij}
\nonumber\\
\end{eqnarray}
We finally enforce Einstein's equation $G_{ab} = 8\pi T_{ab}$ and
simplify using the conservation relations (\ref{eq:Sconstraint}) --
(\ref{eq:sigma_i_constraint}); this leads to the following field
equations:
\begin{eqnarray}
\nabla^2\Theta &=& -8\pi\rho\;,
\label{eq:Thetaeqn}\\
\nabla^2\Phi &=& 4\pi\left(\rho + 3P - 3\dot S\right)\;,
\label{eq:Phieqn}\\
\nabla^2\Xi_i &=& -16\pi S_i\;,
\label{eq:Xieqn}\\
\Box h_{ij}^{\rm TT} &=& -16\pi\sigma_{ij}\;.
\label{eq:hijTTeqn}
\end{eqnarray}

Notice that {\bf only the metric components $h_{ij}^{\rm TT}$ obey a
wave-like equation}.  The other variables $\Theta$, $\Phi$ and $\Xi_i$
are determined by Poisson-type equations.  Indeed, in a purely vacuum
spacetime, the field equations reduce to five Laplace equations and a
wave equation:
\begin{eqnarray}
\nabla^2\Theta^{\rm vac} &=& 0\;,
\\
\nabla^2\Phi^{\rm vac} &=& 0\;,
\\
\nabla^2\Xi_i^{\rm vac} &=& 0\;,
\\
\Box h_{ij}^{\rm TT, vac} &=& 0\;.
\end{eqnarray}
This manifestly demonstrates that only the $h_{ij}^{\rm TT}$ metric
components --- the transverse, traceless degrees of freedom of the
metric perturbation --- characterize the radiative degrees of freedom
in the spacetime.  Although it is possible to pick a gauge in which
other metric components {\it appear} to be radiative, they will not
be: Their radiative character is an illusion arising due to the choice
of gauge or coordinates.

The field equations (\ref{eq:Thetaeqn}) -- (\ref{eq:hijTTeqn}) also
demonstrate that, far from a dynamic, radiating source, the
time-varying portion of the physical degrees of freedom in the metric
is dominated by $h_{ij}^{\rm TT}$.  If we expand the gauge invariant
fields $\Phi$, $\Theta$, $\Xi_i$ and $h_{ij}^{\rm TT}$ in powers of
$1/r$, then, at sufficiently large distances, the leading-order
$O(1/r)$ terms will dominate.  For the fields $\Theta$, $\Phi$ and
$\Xi_i$, the coefficients of the $1/r$ pieces are simply the conserved
mass $\int d^3x \rho$ or the conserved linear momentum $- \int d^3 x
S_i$, from the conservation relations (\ref{eq:Sconstraint}) --
(\ref{eq:sigma_i_constraint}).  Thus, the only time-varying piece of
the physical degrees of freedom in the metric perturbation at order
$O(1/r)$ is the TT piece $h_{ij}^{\rm TT}$.  An alternative proof of
this result is given in Exercise 19.1 of Misner, Thorne and Wheeler
{\cite{mtw}}.

Although the variables $\Phi$, $\Theta$, $\Xi_i$ and $h_{ij}^{\rm TT}$
have the advantage of being gauge invariant, they have the
disadvantage of being non-local.  Computation of these variables at a
point requires knowledge of the metric perturbation $h_{ab}$
everywhere.  This non-locality obscures the fact that the physical,
non-radiative degrees of freedom are causal, a fact which is explicit
in Lorentz gauge
\footnote{One way to see that the guage invariant
degrees of freedom are causal is to combine the vacuum wave equation
(\protect{\ref{eq:elin1}}) for the metric perturbation with the
expression (\protect{\ref{eq:riemann}}) for the gauge-invariant
Riemann tensor.  This gives the wave equation $\Box
R_{\alpha\beta\gamma\delta}=0$.}.  On the other hand, many
observations that seek to 
detect GWs are sensitive only to the value of the Riemann tensor at a
given point in space (see Sec.\ \ref{sec:detect}).  For example, the
Riemann tensor components $R_{itjt}$, which are directly observable by
detectors such as LIGO, are given in terms of the gauge invariant
variables as
\begin{equation}
R_{itjt} = - \frac{1}{2} {\ddot h}_{ij}^{\rm TT} + \Phi_{,ij} + {\dot
  \Xi}_{(i,j)} - \frac{1}{2} {\ddot \Theta} \delta_{ij}.
\label{eq:Riemanngeneral}
\end{equation}
Thus, at least certain combinations of the gauge invariant variables
are locally observable.

\subsection{Local regions of spacetime}
\label{subsec:local}

In the previous subsection we described a splitting of metric
perturbations into radiative, non-radiative, and gauge pieces.  This
splitting requires that the linearized Einstein equations be valid
throughout the spacetime.  However, this assumption is not valid in
the real Universe: Many sources of GWs are intrinsically strong field
sources and cannot be described using linearized theory, and on
cosmological scales the metric of our Universe is not close to the
Minkowski metric.  Furthermore the splitting requires a knowledge of
the metric throughout all of spacetime, whereas any measurements or
observations can probe only finite regions of spacetime.  For these
reasons it is useful to consider linearized perturbation theory in
finite regions of spacetime, and to try to define gravitational
radiation in this more general context.

Consider therefore a finite volume ${\cal V}$ in space.  Can we split
up the metric perturbation $h_{ab}$ in ${\cal V}$ into radiative and
non-radiative pieces?  In general, the answer is no: Within any finite
region GWs cannot be distinguished from time-varying near-zone fields
generated by sources outside that region.  One way to see this is to
note that in finite regions of space, the decomposition of the metric
into various pieces becomes non-unique, as does the decomposition of
vectors into transverse and longitudinal pieces.  [For example the
vector $(x^2 - y^2) \partial_z$ is both transverse and longitudinal.]
Alternatively, we note that within any finite vacuum region ${\cal
V}$, one can {\it always} find a gauge which is locally TT, that is, a
gauge which satisfies the conditions (\ref{eq:spatial}) --
(\ref{eq:transverse}) within the region.  (This fact does not seem to
be widely known, so we give a proof in \ref{sec:localTT}).  In
particular, this applies to the static Coulomb-type field of a point
source, as long as the source itself is outside of ${\cal V}$.
Consequently, isolating the TT piece of the metric perturbation does
not yield just the radiative degrees of freedom within a local region
-- a TT metric perturbation may also contain, for example,
Coulomb-type fields.

Within finite regions of space, therefore, GWs cannot be defined at a
fundamental level -- one simply has time-varying gravitational fields.
However, there is a certain limit in which GWs can be approximately
defined in local regions, namely the limit in which the wavelength of
the waves is much smaller than length and time scales characterizing
the background metric.  This definition of gravitational radiation is
discussed in detail and in a more general context in Sec.\
\ref{sec:lin_in_curved}.  As discussed in that section, this limit
will always be valid when one is sufficiently far from all radiating
sources.

\section{Interaction of gravitational waves with a detector}
\label{sec:detect}

The usual notion of ``gravitational force'' disappears in general
relativity, replaced instead by the idea that freely falling bodies
follow geodesics in spacetime.  Given a spacetime metric $g_{ab}$ and
a set of spacetime coordinates $x^a$, geodesic trajectories are given
by the equation
\begin{equation}
\frac{d^2 x^a}{d\tau^2} +
{\Gamma^a}_{bc}\frac{dx^b}{d\tau}\frac{dx^c}{d\tau} = 0\;,
\label{eq:geod_eqn}
\end{equation}
where $\tau$ is proper time as measured by an observer travelling
along the geodesic.  By writing the derivatives in the geodesic
equation (\ref{eq:geod_eqn}) in terms of coordinate time $t$ rather
than proper time $\tau$, and by combining the $a=t$ equation with the
spatial, $a=j$ equations, we obtain an equation for the coordinate
acceleration:
\begin{equation}
\frac{d^2 x^i}{dt^2} = - (\Gamma^i_{tt} + 2 \Gamma^i_{tj} v^j +
\Gamma^i_{jk} v^j v^k) + v^i (\Gamma^t_{tt} + 2 \Gamma^t_{tj} v^j +
\Gamma^t_{jk} v^j v^k),
\label{eq:geod_eqn1}
\end{equation}
where $v^i = dx^i/dt$ is the coordinate velocity.

Let us now specialize to linearized theory, with the non-flat part of
our metric dominated by a GW in TT gauge.  Further, let us specialize
to non-relativistic motion for our test body.  This implies that $v^i
\ll 1$, and to a good approximation we can neglect the velocity
dependent terms in Eq.\ (\ref{eq:geod_eqn1}):
\begin{equation}
\frac{d^2 x^i}{dt^2} + {\Gamma^i}_{tt} = 0\;.
\end{equation}
In linearized theory and TT gauge,
\begin{equation}
{\Gamma^i}_{tt} = \Gamma_{itt} = \frac{1}{2}\left(2\partial_t h^{\rm
TT}_{jt} - \partial_j h^{\rm TT}_{tt}\right) = 0
\label{eq:linGamma}
\end{equation}
since $h^{\rm TT}_{at} = 0$.  Hence, we find that $d^2x^i/dt^2 = 0$.

Does this result mean that the GW has no effect?  Certainly not!  It
just tells us that in TT gauge the {\it coordinate location} of a
slowly moving, freely falling body is unaffected by the GW.  In
essence, the coordinates move with the waves.

This result illustrates why, in general relativity, it is important to
focus upon coordinate-invariant observables --- a naive interpretation
of the above result would be that freely falling bodies are not
influenced by GWs.  In fact the GWs cause the {\it proper separation}
between two freely falling particles to oscillate, even if the {\it
coordinate separation} is constant.  Consider two spatial freely
falling particles, located at $z = 0$, and separated on the $x$ axis
by a coordinate distance $L_c$.  Consider a GW in TT gauge that
propagates down the $z$ axis, $h^{\rm TT}_{ab}(t,z)$.  The proper
distance $L$ between the two particles in the presence of the GW is
given by
\begin{eqnarray}
L &=& \int_0^{L_c} dx\,\sqrt{g_{xx}} = \int_0^{L_c} dx\,\sqrt{1 +
h^{\rm TT}_{xx}(t,z = 0)}
\nonumber\\
&\simeq& \int_0^{L_c} dx\,\left[1 + \frac{1}{2} h^{\rm TT}_{xx}(t,z =
  0)\right] = L_c\left[1 + \frac{1}{2} h^{\rm TT}_{xx}(t,z =
  0)\right]\;.
\label{eq:waveeffect}
\end{eqnarray}
Notice that we use the fact that the coordinate location of each
particle is fixed in TT gauge!  In a gauge in which the particles move
with respect to the coordinates, the limits of integration would have
to vary.  Equation (\ref{eq:waveeffect}) tells us that the proper
separation of the two particles oscillates with a fractional length
change $\delta L/L$ given by
\begin{equation}
\frac{\delta L}{L} \simeq \frac{1}{2} h^{\rm TT}_{xx}(t,z = 0)\;.
\label{eq:strainans}
\end{equation}

Although we used TT gauge to perform this calculation, the result is
gauge independent; we will derive it in a different gauge momentarily.
Notice that $h^{\rm TT}_{xx}$ acts as a strain --- a fractional length
change.  The magnitude $h$ of a wave is often referred to as the
``wave strain''.  The proper distance we have calculated here is a
particularly important quantity since it directly relates to the
accumulated phase which is measured by laser interferometric GW
observatories (cf.\ the contribution by Danzmann in this volume).  The
``extra'' phase $\delta \phi$ accumulated by a photon that travels
down and back the arm of a laser interferometer in the presence of a
GW is $\delta\phi = 4\pi \delta L/\lambda$, where $\lambda$ is the
photon's wavelength and $\delta L$ is the distance the end mirror
moves relative to the beam splitter\footnote{This description of the
phase shift only holds if $L \ll \lambda$, so that the metric
perturbation does not change value very much during a light travel
time.  This condition will be violated in the high frequency regime
for space-based GW detectors; a more careful analysis of the phase
shift is needed in this case {\cite{lhh00}}.}.  We now give a
different derivation of the fractional length change
(\ref{eq:strainans}) based on the concept of {\it geodesic deviation}.
Consider a geodesic in spacetime given by $x^a = z^a(\tau)$, where
$\tau$ is the proper time, with four velocity $u^a(\tau) =
dz^a/d\tau$.  Suppose we have a nearby geodesic $x^a(\tau) = z^a(\tau)
+ L^a(\tau)$, where $L^a(\tau)$ is small.  We can regard the
coordinate displacement $L^a$ as a vector ${\vec L} = L^a \partial_a$
on the first geodesic; this is valid to first order in ${\vec L}$.
Without loss of generality, we can make the connecting vector be
purely spatial: $L^a u_a =0$.  Spacetime curvature causes the
separation vector to change with time --- the geodesics will move
further apart or closer together, with an acceleration given by the
geodesic deviation equation
\begin{equation}
u^b \nabla_b (u^c \nabla_c L^a) = - {R^a}_{bcd}[{\vec z}(\tau)] u^b
L^c u^d\;;
\label{eq:geod_dev}
\end{equation}
see, e.g., Ref.\ {\cite{jh03}}, Chap.\ 21.  This equation is valid to
linear order in $L^a$; fractional corrections to this equation will
scale as $L / {\cal L}$, where ${\cal L}$ is the lengthscale over
which the curvature varies.

For application to GW detectors, the shortest such lengthscale ${\cal
L}$ is the wavelength $\lambda$ of the GWs.  Thus, the geodesic
deviation equation will have fractional corrections of order $L /
\lambda$.  For ground-based detectors $L \lesssim $ a few km, while
$\lambda \gtrsim 3000 {\rm km}$ (see Sec.\ \ref{subsec:high}); thus
the approximation will be valid.  For detectors with $L \gtrsim
\lambda$ (e.g. the space based detector LISA) the analysis here is not
valid and other techniques must be used to analyze the detector.

A convenient coordinate system for analyzing the geodesic deviation
equation (\ref{eq:geod_dev}) is the {\it local proper reference frame}
of the observer who travels along the first geodesic.  This coordinate
system is defined by the requirements
\begin{equation}
z^i(\tau) = 0,\ \ \ \ \ g_{ab}(t,{\bf 0}) = \eta_{ab}, \ \ \ \
\Gamma^a_{bc}(t,{\bf 0}) =0,
\label{eq:lprf}
\end{equation}
which imply that the metric has the form
\begin{equation}
ds^2 = - dt^2 + d {\bf x}^2 + O\left(\frac{{\bf x}^2}{{\cal R}^2}
\right).
\label{eq:lprfmetric}
\end{equation}
Here ${\cal R}$ is the radius of curvature of spacetime, given by
${\cal R}^{-2} \sim ||R_{abcd}||$.  It also follows from the gauge
conditions (\ref{eq:lprf}) that proper time $\tau$ and coordinate time
$t$ coincide along the first geodesic, that ${\vec u} = \partial_t$
and that $L^a =(0,L^i)$.

Consider now the proper distance between the two geodesics, which are
located at $x^i=0$ and $x^i = L^i(t)$.  From the metric
(\ref{eq:lprfmetric}) we see that this proper distance is just $|{\bf
L}| = \sqrt{L_i L_i}$, up to fractional corrections of order $L^2 /
{\cal R}^2$.  For a GW of amplitude $h$ and wavelength $\lambda$ we
have ${\cal R}^{-2} \sim h / \lambda^2$, so the fractional errors are
$\sim h L^2 / \lambda^2$.  (Notice that ${\cal R} \sim {\cal
L}/\sqrt{h}$ --- the wave's curvature scale ${\cal R}$ is much larger
than the lengthscale ${\cal L}$ characterizing its variations.)  Since we are
restricting attention to detectors with $L \ll \lambda$, these
fractional errors are much smaller than the fractional distance
changes $\sim h$ caused by the GW [Eq.\ (\ref{eq:strainans})].
Therefore, we can simply identify $|{\bf L}|$ as the proper
separation.

We now evaluate the geodesic deviation equation (\ref{eq:geod_dev})
in the local proper reference frame coordinates.  From the conditions
(\ref{eq:lprf}) it follows that we can replace the covariant time
derivative operator $u^a \nabla_a$ with $\partial / (\partial t)$.
Using ${\vec u} = \partial_t$ and $L^a = (0,L^i)$ we get
\begin{equation}
\frac{d^2 L^i(t)}{dt^2} =-  {R}_{itjt}(t,{\bf 0}) L^j(t) \;.
\label{eq:observable}
\end{equation}
Note that the key quantity entering into the equation, $R_{itjt}$, is
gauge invariant in linearized theory, so we can use any convenient
coordinate system to evaluate it.  Using the expression
(\ref{eq:Rtitj}) for the Riemann tensor in terms of the TT gauge
metric perturbation $h_{ij}^{\rm TT}$ we find
\begin{equation}
\frac{d^2 L^i}{dt^2} = \frac{1}{2}\frac{d^2h^{\rm TT}_{ij}}{dt^2}L^j\;.
\label{eq:geod_dev_detector}
\end{equation}
Integrating this equation using $L^i(t) = L^i_0 + \delta L^i(t)$ with
$|\delta {\bf L}| \ll |{\bf L}_0|$ gives
\begin{equation}
\delta L^i(t) = \frac{1}{2}h^{\rm TT}_{ij}(t) L_0^j\;.
\label{eq:response}
\end{equation}

This equation is ideal for analyzing an interferometric GW detector.
We choose Cartesian coordinates such that the interferometer's two
arms lie along the $x$ and $y$ axes, with the beam splitter at the
origin.  For concreteness, let us imagine that the GW propagates down
the $z$-axis.  Then, as discussed in Sec.\ \ref{sec:TTgauge}, the only
non-zero components of the metric perturbation are $h^{\rm TT}_{xx} =
-h^{\rm TT}_{yy} = h_+$ and $h^{\rm TT}_{xy} = h^{\rm TT}_{yx} =
h_\times$, where $h_+(t-z)$ and $h_\times(t-z)$ are the two
polarization components.  We take the ends of one of the
interferometer's two arms as defining the two nearby geodesics; the
first geodesic is defined by the beam splitter at ${\bf x}=0$, the
second by the end-mirror.  From Eq.\ (\ref{eq:response}) we then find
that the distances $L = | {\bf L}|$ of the arms' ends from the beam
splitter vary with time as
\begin{eqnarray}
\frac{\delta L_x}{L} &=& \frac{1}{2} h_+\;,
\nonumber\\
\frac{\delta L_y}{L} &=& -\frac{1}{2} h_+\;.
\end{eqnarray}
(Here the subscripts $x$ and $y$ denote the two different arms, not
the components of a vector).  These distance changes are then measured
via laser interferometry.  Notice that the GW (which is typically a
sinusoidally varying function) acts tidally, squeezing along one axis
and stretching along the other.  In this configuration the detector is
sensitive only to the $+$ polarization of the GW.  The $\times$
polarization acts similarly, except that it squeezes and stretches
along a set of axes that are rotated with respect to the $x$ and $y$
axes by $45^\circ$.  The force lines corresponding to the two
different polarizations are illustrated in Fig.\
{\ref{fig:forcelines}}.

\begin{figure}[t]
\includegraphics[width = 13cm]{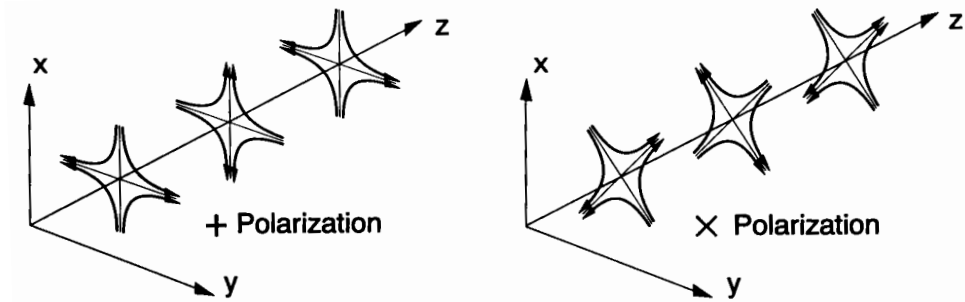}
\caption{Lines of force for a purely $+$ GW (left), and for a purely
$\times$ GW (right).  Figure kindly provided by Kip Thorne; originally
published in Ref.\ {\cite{science92}}.}
\label{fig:forcelines}
\end{figure}

Of course, we don't expect nature to provide GWs that so perfectly
align with our detectors.  In general, we will need to account for the
detector's {\it antenna pattern}, meaning that we will be sensitive to
some weighted combination of the two polarizations, with the weights
depending upon the location of a source on the sky, and the relative
orientation of the source and the detector.  See Ref.\
{\cite{300yrs}}, Eqs.\ (104a,b) and associated text for further
discussion.

Finally, in our analysis so far of detection we have assumed that the
only contribution to the metric perturbation is the GW contribution.
However, in reality time-varying near zone gravitational fields
produced by sources in the vicinity of the detector will also be
present.  From Eq.\ (\ref{eq:observable}) we see that the quantity
that is actually measured by interferometric detectors is the
space-time-space-time or electric-type piece $R_{itjt}$ of the Riemann
tensor (or more precisely the time-varying piece of this within the
frequency band of the detector).  From the general expression
(\ref{eq:Riemanngeneral}) for this quantity we see that $R_{itjt}$
contains contributions from both $h_{ij}^{\rm TT}$ describing GWs, and
also additional terms describing the time-varying near zone
gravitational fields.  There is no way for the detector to separate
these two contributions, and the time-varying near zone gravitational
fields produced by motions of bedrock, air, human bodies, and
tumbleweeds can all contribute to the output of the detector and act
as sources of noise \cite{ht1998,tw1999,tcreighton}.

\section{The generation of gravitational waves: Putting in the source}
\label{sec:lin_with_source}

\subsection{Slow-motion sources in linearized gravity}

Gravitational waves are generated by the matter source term on the
right hand side of the linearized Einstein equation
\begin{equation}
\Box {\bar h}_{ab} = -16\pi T_{ab}\;,
\label{eq:LinEinstein_TT}
\end{equation}
cf.\ Eq.\ (\ref{eq:elin}) (presented here in Lorentz gauge).  In this
section we will compute the leading order contribution to the spatial
components of the metric perturbation for a source whose internal
motions are slow compared to the speed of light (``slow-motion
sources'').  We will then compute the TT piece of the metric
perturbation to obtain the standard quadrupole formula for the emitted
radiation.

Equation (\ref{eq:LinEinstein_TT}) can be solved by using a Green's
function.  A wave equation with source generically takes the form
\begin{equation}
\Box f(t, {\bf x}) = s(t, {\bf x})\;,
\label{eq:wave_source}
\end{equation}
where $f(t,{\bf x})$ is the radiative field, depending on time $t$ and
position ${\bf x}$, and $s(t,{\bf x})$ is a source function.  The
Green's function $G(t,{\bf x};t',{\bf x}')$ is the field which arises
due to a delta function source; it tells how much field is generated
at the ``field point'' $(t,{\bf x})$ per unit source at the ``source
point'' $(t',{\bf x}')$:
\begin{equation}
\Box G(t, {\bf x}; t',{\bf x}') = \delta(t - t')\delta({\bf x} - {\bf x}')\;.
\label{eq:wave_greens}
\end{equation}
The field which arises from our actual source is then given by
integrating the Green's function against $s(t,{\bf x})$:
\begin{equation}
f(t, {\bf x}) = \int dt'd^3 x'\,G(t, {\bf x}; t',{\bf x}')\,s(t',{\bf
x}')\;\;.
\end{equation}
The Green's function associated with the wave operator $\Box$ is very
well known (see, e.g., {\cite{jackson}}):
\begin{equation}
G(t, {\bf x}; t', {\bf x}') = -\frac{\delta(t' - [t - |{\bf x} - {\bf
x}'|/c])}{4\pi|{\bf x} - {\bf x}'|}\;.
\label{eq:BoxGreen}
\end{equation}
The quantity $t - |{\bf x} - {\bf x}'|/c$ is the {\it retarded time};
it takes into account the lag associated with the propagation of
information from events at ${\bf x}$ to position ${\bf x}'$.  The
speed of light $c$ has been restored here to emphasize the causal
nature of this Green's function; we set it back to unity in what
follows.

Applying this result to Eq.\ ({\ref{eq:LinEinstein_TT}}), we find
\begin{equation}
{\bar h}_{ab}(t,{\bf x}) = 4\int d^3x' \frac{T_{ab}(t - |{\bf x} - {\bf x}'|,
{\bf x}')}{|{\bf x} - {\bf x}'|}\;.
\label{eq:naivesolution}
\end{equation}
As already mentioned, the radiative degrees of freedom are contained
entirely in the spatial part of the metric, projected transverse and
traceless.  First, consider the spatial part of the metric:
\begin{equation}
{\bar h}_{ij}(t,{\bf x}) = 4\int d^3x' \frac{T^{ij}(t - |{\bf x} - {\bf
x}'|,{\bf x}')} {|{\bf x} - {\bf x}'|}\;.
\label{eq:spatialsolution}
\end{equation}
We have raised indices on the right-hand side, using the rule that the
position of spatial indices in linearized theory is irrelevant.

We now evaluate this quantity at large distances from the source.
This allows us to replace the factor $|{\bf x} - {\bf x}'|$ in the
denominator with $r = |{\bf x}|$.  The corresponding fractional errors
scale as $\sim L / r$, where $L$ is the size of the source; these
errors can be neglected.  We also make the same replacement in the
time argument of $T_{ij}$:
\begin{equation}
T_{ij}(t-|{\bf x} - {\bf x}'|,{\bf x}') \approx T_{ij}(t-r,{\bf x}').
\label{eq:slowmotion}
\end{equation}
Using the formula $|{\bf x} - {\bf x}'| = r - n^i x^{'\,i} + O(1/r)$
where $n^i = x^i/r$, we see that the fractional errors generated by
the replacement (\ref{eq:slowmotion}) scale as $L/\tau$, where $\tau$
is the timescale over which the system is changing.  This quantity is
just the velocity of internal motions of the source (in units with
$c=1$), and is therefore small compared to one by our assumption.
These replacements give
\begin{equation}
{\bar h}_{ij}(t,{\bf x}) = \frac{4}{r}\int d^3x'\,T^{ij}(t - r,{\bf x}')\;,
\label{eq:almost_quadrupole}
\end{equation}
which is the first term in a multipolar expansion of the radiation field.

Equation (\ref{eq:almost_quadrupole}) almost gives us the quadrupole
formula that describes GW emission (at leading order).  To get the
rest of the way there, we need to massage this equation a bit.  The
stress-energy tensor must be conserved, which means $\partial_a T^{ab}
= 0$ in linearized theory.  Breaking this up into time and space
components, we have
\begin{eqnarray}
\partial_t T^{tt} + \partial_i T^{ti} &=& 0\,,
\\
\partial_t T^{ti} + \partial_j T^{ij} &=& 0\,.
\label{eq:cons_of_stress1}
\end{eqnarray}
From this, it follows rather simply that
\begin{equation}
\partial^2_t T^{tt} = \partial_k\partial_l T^{kl}\;.
\label{eq:cons_of_stress2}
\end{equation}
Multiply both sides of this equation by $x^i x^j$.  We first
manipulate the left-hand side:
\begin{equation}
\partial^2_t T^{tt} x^i x^j = \partial_t^2\left(T^{tt} x^i x^j\right)\;.
\end{equation}
Next, manipulate the right-hand side of Eq.\ (\ref{eq:cons_of_stress2}),
multiplied by $x^i x^j$:
\begin{equation}
\partial_k\partial_l T^{kl} x^i x^j =
\partial_k\partial_l \left(T^{kl} x^i x^j\right) -
2\partial_k\left(T^{ik} x^j + T^{kj} x^i\right) + 2 T^{ij}\;.
\end{equation}
This identity is easily verified\footnote{Although one of us (SAH) was
unable to do this simple calculation while delivering lectures at a
summer school in Brownsville, TX.  Never attempt to derive the
quadrupole formula while medicated.} by expanding the derivatives and
applying the identity $\partial_i x^j = {\delta_i}^j$.  We thus have
\begin{equation}
\partial_t^2\left(T^{tt} x^i x^j\right) =
\partial_k\partial_l \left(T^{kl} x^i x^j\right) -
2\partial_k\left(T^{ik} x^j + T^{kj} x^i\right) + 2 T^{ij}\;.
\end{equation}
This yields
\begin{eqnarray}
\frac{4}{r} \int d^3 x'\,T_{ij} &=&
\frac{4}{r}\int d^3x'
\left[\frac{1}{2}\partial_t^2\left(T^{tt} x^{\prime i} x^{\prime
j}\right) + \partial_k\left(T^{ik} x^{\prime j} + T^{kj} x^{\prime
i}\right)\right.\nonumber\\
& &\left.\qquad\qquad - \frac{1}{2} \partial_k\partial_l \left(T^{kl}
x^{\prime i}x^{\prime j}\right)\right]
\nonumber\\
&=& \frac{2}{r}\int d^3x'\,\partial_t^2\left(T^{tt} x^{\prime i}
x^{\prime j}\right)
\nonumber\\
&=& \frac{2}{r}\frac{\partial^2}{\partial t^2}\int d^3x'\,T^{tt}
x^{\prime i} x^{\prime j}
\nonumber\\
&=& \frac{2}{r}\frac{\partial^2}{\partial t^2}\int
d^3x'\,\rho\,x^{\prime i} x^{\prime j}\;.
\label{eq:oh_so_close}
\end{eqnarray}
In going from the first line to the second, we used the fact that the
second and third terms under the integral are divergences.  Using
Gauss's theorem, they can thus be recast as surface integrals; taking
the surface outside the source, their contribution is zero.  In going
from the second line to the third, we used the fact that the
integration domain is not time dependent, so we can take the
derivatives out of the integral.  Finally, we used the fact that
$T^{tt}$ is the mass density $\rho$.  Defining the second moment
$I_{ij}$ of the mass distribution via
\begin{equation}
I_{ij}(t) = \int d^3x'\,\rho(t,{\bf x}') \,x^{\prime i}x^{\prime j}\;,
\label{eq:quad_moment1}
\end{equation}
and combining Eqs.\ (\ref{eq:almost_quadrupole}) and
(\ref{eq:oh_so_close}) now gives
\begin{equation}
{\bar h}_{ij}(t,{\bf x}) = \frac{2}{r}\frac{d^2I_{ij}(t-r)}{dt^2}\;.
\label{eq:h_from_I}
\end{equation}
When we subtract the trace from $I_{ij}$, we obtain the {\it
quadrupole moment} tensor:
\begin{equation}
{\cal I}_{ij} = I_{ij} - \frac{1}{3}\delta_{ij} I, \qquad
I = I_{ii}\;.
\label{eq:quad_moment}
\end{equation}
This tensor will prove handy shortly.

To complete the derivation, we must project out the non-TT pieces of
the right-hand side of Eq.\ (\ref{eq:h_from_I}).
Since we are working to leading order in $1/r$, at each field point
${\bf x}$ this operation reduces to algebraically projecting the
tensor perpendicularly to the local direction of propagation ${\bf n}
= {\bf x} / r$, and subtracting off the trace.
It is useful to introduce the projection tensor,
\begin{equation}
P_{ij} = \delta_{ij} - n_i n_j\;.
\end{equation}
This tensor eliminates vector components parallel to ${\bf n}$,
leaving only transverse components.  Thus,
\begin{equation}
{\bar h}^T_{ij} = {\bar h}_{kl}P_{ik}P_{jl}
\end{equation}
is a transverse tensor.  Finally, we remove the trace; what remains is
\begin{eqnarray}
h^{\rm TT}_{ij} &=& {\bar h}_{kl} P_{ik} P_{jl} -
\frac{1}{2}P_{ij}P_{kl}{\bar h}_{kl}\;.
\label{eq:quad_formula1}
\end{eqnarray}
Substituting Eq.\ (\ref{eq:h_from_I}) into (\ref{eq:quad_formula1}),
we obtain our final quadrupole formula:
\begin{equation}
h^{\rm TT}_{ij}(t,{\bf x}) = \frac{2}{r}\frac{d^2{\cal
I}_{kl}(t-r)}{dt^2}\left[P_{ik}({\bf n})P_{jl}({\bf n}) -
\frac{1}{2}P_{kl}({\bf n})P_{ij}({\bf n})\right] \;.
\label{eq:quad_formula}
\end{equation}

\subsection{Extension to sources with non-negligible self gravity}
\label{sec:quad1}

Our derivation of the quadrupole formula (\ref{eq:quad_formula}) assumed
the validity of the linearized Einstein equations.  In particular, the
derivation is not applicable to systems with weak (Newtonian) gravity
whose dynamics are dominated by self-gravity, such as binary star
systems\footnote{Stress energy conservation in linearized gravity,
$\partial^a T_{ab} =0$, forces all bodies to move on geodesics of
the Minkowski metric.}.  This shortcoming of the above
linearized-gravity derivation of the quadrupole formula was first
pointed out by Eddington.  However, it is very straightforward to
extend the derivation to encompass systems with non-negligible self gravity.

In full general relativity we define the quantity ${\bar h}^{ab}$ via
\begin{equation}
\sqrt{-g} g^{ab} = \eta^{ab} - {\bar h}^{ab},
\end{equation}
where $\eta^{ab} \equiv {\rm diag}(-1,1,1,1)$.
When gravity is weak this definition coincides with
our previous definition of ${\bar h}^{ab}$ as a trace-reversed metric
perturbation.  We impose the harmonic gauge condition
\begin{equation}
\partial_a (\sqrt{-g} g^{ab}) = \partial_a {\bar h}^{ab} =0.
\label{eq:harmonic}
\end{equation}
In this gauge the Einstein equation can be written
\begin{equation}
\Box_{\rm flat} {\bar h}^{ab} = - 16 \pi ( T^{ab} + t^{ab} )
\label{eq:enonlin}
\end{equation}
where $\Box_{\rm flat} \equiv \eta^{ab} \partial_a \partial_b$ is the
flat-spacetime wave operator, and $t^{ab}$ is a pseudotensor that is
constructed from ${\bar h}^{ab}$. Taking a coordinate divergence of
this equation and using the gauge condition (\ref{eq:harmonic}) shows
that stress-energy conservation can be written
\begin{equation}
\partial_a (T^{ab} + t^{ab}) =0.
\label{eq:nonlincons}
\end{equation}

Equations (\ref{eq:harmonic}), (\ref{eq:enonlin}) and
(\ref{eq:nonlincons}) are precisely the same equations as are used in
the linearized-gravity derivation of the quadrupole formula, except
for the fact that the stress energy tensor $T^{ab}$ is replaced by
$T^{ab} + t^{ab}$.  Therefore the derivation of the last subsection
carries over, with the one modification that the formula
(\ref{eq:quad_moment1}) for $I_{ij}$ is replaced by
\begin{equation}
I_{ij}(t) = \int d^3 x' \left[T^{tt}(t,{\bf x}') + t^{tt}(t,{\bf
    x}')\right] x^{\prime i}x^{\prime j}.
\end{equation}
In this equation the term $t^{tt}$ describes gravitational binding
energy, roughly speaking. For systems with weak gravity, this term is
negligible in comparison with the term $T^{tt}$ describing the
rest-masses of the bodies.  Therefore the quadrupole formula
(\ref{eq:quad_formula}) and the original definition
(\ref{eq:quad_moment1}) of $I_{ij}$ continue to apply to the more
general situation considered here.

\subsection{Dimensional analysis}

The rough form of the leading GW field that we just derived, Eq.\
(\ref{eq:quad_formula}), can be deduced using simple physical
arguments.  First, we define some moments of the mass distribution.
The zeroth moment is just the mass itself:
\begin{equation}
M_0 \equiv \int \rho\,d^3x = M\;.
\end{equation}
(More accurately, this is the total mass-energy of the source.)
Next, we define the dipole moment:
\begin{equation}
M_1 \equiv \int \rho\,x_i\,d^3x = ML_i\;.
\end{equation}
$L_i$ is a vector with the dimension of length; it describes the
displacement of the center of mass from our chosen origin.  (As such,
$M_1$ is clearly not a very meaningful quantity --- we can change its
value simply by choosing a different origin.)

If our mass distribution exhibits internal motion, then moments of the
{\it mass current}, $j_i = \rho v_i$, are also important.  The first
moment is the spin angular momentum:
\begin{equation}
S_1 \equiv \int \rho v_j\,x_k\,\epsilon_{ijk}\,d^3x = S_i\;.
\end{equation}
Finally, we look at the second moment of the mass distribution:
\begin{equation}
M_2 \equiv \int \rho\,x_i\,x_j\,d^3x = M L_{ij}
\end{equation}
where $L_{ij}$ is a tensor with the dimension length squared.

Using dimensional analysis and simple physical arguments, it is simple
to see that the first moment that can contribute to GW emission is
$M_2$.  Consider first $M_0$.  We want to combine $M_0$ with the
distance to our source, $r$, in such a way as to produce a
dimensionless wavestrain $h$.  The only way to do this (bearing in
mind that the strain should fall off as $1/r$, and restoring factors
of $G$ and $c$) is to put
\begin{equation}
h \sim \frac{G}{c^2}\frac{M_0}{r}\;.
\label{eq:hnewton}
\end{equation}
Does this formula make sense for radiation?  Not at all!  Conservation
of mass-energy tells us that $M_0$ for an isolated source cannot vary
dynamically.  This $h$ cannot be radiative; it corresponds to a
Newtonian potential, rather than a GW.

How about the moment $M_1$?  In order to get the dimensions right, we
must take one time derivative:
\begin{equation}
h \sim \frac{G}{c^3}\frac{d}{dt}\frac{M_1}{r}\;.
\end{equation}
(The extra factor of $c$ converts the dimension of the time derivative
to space, so that the whole expression is dimensionless.)  Think
carefully about the derivative of $M_1$:
\begin{equation}
\frac{dM_1}{dt} = \frac{d}{dt}\int\rho\,x_i\,d^3x =
\int\rho\,v_i\,d^3x = P_i\;.
\end{equation}
This is the total momentum of our source.  Our guess for the form of a
wave corresponding to $M_1$ becomes
\begin{equation}
h \sim \frac{G}{c^3}\frac{P}{r}\;.
\label{eq:hboost}
\end{equation}
Can this describe a GW?  Again, not a chance: The momentum of an
isolated source must be conserved.  By boosting into a different
Lorentz frame, we can always set $P = 0$.  Terms like this can only be
gauge artifacts; they do not correspond to radiation.  [Indeed, terms
like (\ref{eq:hboost}) appear in the metric of a moving black hole,
and correspond to the relative velocity of the hole and the observer;
see {\cite{membrane}}, Chapter 5.]

How about $S_1$?  Dimensional analysis tells us that radiation from
$S_1$ must take the form
\begin{equation}
h \sim \frac{G}{c^4}\frac{d}{dt}\frac{S_1}{r}.
%\quad\mbox{?}
\end{equation}
Conservation of angular momentum tells us that the total spin of an
isolated system cannot change, so we reject this term for the same
reason that we rejected (\ref{eq:hnewton}) --- it cannot correspond to
radiation.

Finally, we examine $M_2$:
\begin{equation}
h \sim \frac{G}{c^4}\frac{d^2}{dt^2}\frac{M_2}{r}\;.
\end{equation}
There is {\it no} conservation principle that allows us to reject this
term.  Comparing to Eq.\ (\ref{eq:quad_formula}), we see that this is
the quadrupole formula we derived earlier, up to numerical factors.

In ``normal'' units, the prefactor of this formula turns out to be
$G/c^4$ --- a small number divided by a {\it very} big number.  In
order to generate interesting amounts of GWs, the quadrupole moment's
variation must be enormous.  The only interesting sources of GWs will
be those which have very large masses undergoing extremely rapid
variation; even in this case the strain we expect from typical sources
is tiny.  The smallness of GWs reflects the fact that gravity is the
weakest of the fundamental interactions.

\subsection{Numerical estimates}
\label{subsec:numerstrain}

Consider a binary star system, with stars of mass $m_1$ and $m_2$ in a
circular orbit with separation $R$.  The quadrupole moment is given by
\begin{equation}
{\cal I}_{ij} = \mu\left(x_i x_j - \frac{1}{3}R^2\delta_{ij}\right)\;,
\end{equation}
where $\mu = m_1 m_2/(m_1 + m_2)$ is the binary's reduced mass and
${\bf x}$ is the relative displacement, with $|{\bf x}| = R$.  We
use the center-of-mass reference frame, and 
choose the coordinate axes so that the binary lies in the $xy$ plane,
so $x = x_1 = R\cos\Omega t$, $y = x_2 = R\sin\Omega t$, $z = x_3 =
0$.  Let us further choose to evaluate the field on the $z$ axis, so
that ${\bf n}$ points in the $z$-direction.  The projection operators
in Eq.\ (\ref{eq:quad_formula}) then simply serve to remove the $zj$
components of the tensor.  Bearing this in mind, the quadrupole
formula (\ref{eq:quad_formula}) yields
\begin{equation}
h^{\rm TT}_{ij} = \frac{2 \ddot {\cal I}_{ij}}{r}\;.
\end{equation}
The quadrupole moment tensor is
\begin{equation}
{\cal I}_{ij} = \mu R^2
\left[\matrix{
\cos^2\Omega t - \frac{1}{3} & \cos\Omega t\sin\Omega t & 0 \cr
\cos\Omega t\sin\Omega t & \cos^2\Omega t - \frac{1}{3} & 0 \cr
0 & 0 & -\frac{1}{3} \cr}\right]\;;
\end{equation}
its second derivative is
\begin{equation}
\ddot{\cal I}_{ij} = -2\Omega^2\mu R^2
\left[\matrix{
 \cos2\Omega t & \sin2\Omega t & 0 \cr
-\sin2\Omega t &-\cos2\Omega t & 0 \cr
0 & 0 & 0 \cr}\right]\;.
\end{equation}
The magnitude $h$ of a typical non-zero component of $h^{\rm TT}_{ij}$
is
\begin{equation}
h = \frac{4\mu\Omega^2 R^2}{r} = \frac{4\mu M^{2/3}\Omega^{2/3}}{r}\;.
\end{equation}
We used Kepler's 3rd law\footnote{In units with $G = 1$, and for
circular orbits of radius $R$, $R^3\Omega^2 = M$.} to replace $R$ with
powers of the orbital frequency $\Omega$ and the total mass $M = m_1 +
m_2$.
%The combination of masses appearing here, $\mu M^{2/3}$,
%appears quite often in studies of GW emission from binaries; it
%motivates the definition of the {\it chirp mass}:
%\begin{equation}
%{\cal M} = \mu^{3/5}M^{2/5}\;.
%\end{equation}
For the purpose of our numerical estimate, we will take the members of
the binary to have equal masses, so that $\mu = M/4$:
\begin{equation}
h = \frac{M^{5/3}\Omega^{2/3}}{r}\;.
\end{equation}
Finally, we insert numbers corresponding to plausible sources:
\begin{eqnarray}
h &\simeq& 10^{-21}\left(\frac{M}{2\,M_\odot}\right)^{5/3}
\left(\frac{1\,\mbox{hour}}{P}\right)^{2/3}
\left(\frac{1\,\mbox{kiloparsec}}{r}\right)
\nonumber\\
&\simeq& 10^{-22}\left(\frac{M}{2.8\,M_\odot}\right)^{5/3}
\left(\frac{0.01\,\mbox{sec}}{P}\right)^{2/3}
\left(\frac{100\,\mbox{Megaparsecs}}{r}\right)\;.
\end{eqnarray}
The first line corresponds roughly to the mass, distance and orbital
period ($P = 2\pi/\Omega$) expected for the many close binary white
dwarf systems in our galaxy.  Such binaries are so common that they
are likely to be a confusion limited source of GWs for space-based
detectors, acting in some cases as an effective source of noise.  The
second line contains typical parameter values for binary neutron stars
that are on the verge of spiralling together and merging.  Such waves
are targets for the ground-based detectors that have recently begun
operations.  The {\it tiny} magnitude of these waves illustrates why
detecting GWs is so difficult.

\section{Linearized theory of gravitational waves in a curved background}
\label{sec:lin_in_curved}

At the most fundamental level, GWs can only be defined within the
context of an approximation in which the wavelength of the waves is
much smaller than lengthscales characterizing the background spacetime
in which the waves propagate.  In this section, we discuss
perturbation theory of curved spacetimes, describe the approximation
in which GWs can be defined, and derive the effective stress tensor
which describes the energy content of GWs.  The material in this
section draws on the treatments given in Chapter 35 of Misner, Thorne
and Wheeler {\cite{mtw}}, Sec.\ 7.5 of Wald {\cite{wald1984}}, and the
review articles {\cite{thorne1983,schutz2001}}.

\subsection{Perturbation theory of curved vacuum spacetimes}
\label{sec:pertcurved}

Throughout this section we will for simplicity restrict attention to
vacuum spacetime regions.  We consider a one-parameter family of
solutions of the vacuum Einstein equation, parameterized by
$\varepsilon$, of the form
\begin{equation}
g_{ab} = g_{ab}^{\rm B} + \varepsilon h_{ab} + \varepsilon^2 j_{ab} +
O(\varepsilon^3)\;.
\label{eq:metricexpand}
\end{equation}
Here $g_{ab}^{\rm B}$ is the background metric; it was taken to be the
Minkowski metric in Secs.\ {\ref{sec:basic_lin}},
{\ref{sec:lin_with_source}} and {\ref{subsec:gauge_invar}}.  Here we
allow $g_{ab}^{\rm B}$ to be any vacuum solution of the Einstein
equations.  The quantity $h_{ab}$ is the linear order metric
perturbation, as in the previous sections; $j_{ab}$ is a second order
metric perturbation which will be used in Sec.\ {\ref{sec:Tab}}.  We
can regard $\varepsilon$ as a formal expansion parameter; we set its
value to unity at the end of our calculations.

The derivation of the linearized Einstein equation proceeds as before.
Most of the formulae for linearized perturbations of Minkowski
spacetime continue to apply, with $\eta_{ab}$ replaced by $g^{\rm
B}_{ab}$, and with partial derivatives $\partial_a$ replaced by
covariant derivatives with respect to the background, $\nabla^{\rm
B}_a$.  Some of the formulae acquire extra terms involving coupling to
the background Riemann tensor.

Inserting Eq.\ (\ref{eq:metricexpand}) into the formula for connection
coefficients gives
\begin{eqnarray}
\label{eq:Gammaf}
\Gamma^a_{bc} &=& \frac{1}{2}g^{ad}
\left(\partial_c g_{db} + \partial_b g_{dc} -\partial_d g_{bc}\right)
\\
&=& \frac{1}{2}\left( g^{{\rm B}\,ad} - \varepsilon h^{ad} \right)
 \left(\partial_c g^{\rm B}_{db} + \varepsilon \partial_c h_{db}
+ \partial_b g^{\rm B}_{dc}
\right.\nonumber\\
& &\qquad\qquad\qquad\left.+ \varepsilon \partial_b h_{dc}
-\partial_d g^{\rm B}_{bc} - \varepsilon \partial_d h_{bc} \right)
+ O(\varepsilon^2)
\nonumber \\
&=& \Gamma^{{\rm B}\,a}_{\ \ bc} + \varepsilon \delta \Gamma^a_{bc} +
O(\varepsilon^2)\;.
\label{eq:connectioncurved}
\end{eqnarray}
Here $\Gamma^{{\rm B}\,a}_{\ \ bc}$ are the connection coefficients of
the background metric $g_{ab}^{\rm B}$, and the first order
corrections to the connection coefficients are given by
\begin{eqnarray}
\delta \Gamma^a_{bc} &=&
-\frac{1}{2} h^{ad} g^{\rm B}_{de} \Gamma^{{\rm B}\,e}_{\ \ bc}
+ \frac{1}{2} g^{{\rm B}\,ad}
 \left(\partial_c h_{db}
+ \partial_b h_{dc}
-\partial_d h_{bc} \right)
\nonumber
\\
&=& \frac{1}{2}g^{{\rm B}\,ad}
\left(\nabla^{\rm B}_c h_{db} + \nabla^{\rm B}_b h_{dc} -\nabla^{\rm
  B}_d h_{bc}\right)\;,
\label{eq:Gammaf1}
\end{eqnarray}
where $\nabla^{\rm B}_a$ is the covariant derivative operator
associated with the background metric.  Equation (\ref{eq:Gammaf1})
can be derived more easily, at any given point in spacetime, by
evaluating the expression (\ref{eq:Gammaf}) in a coordinate system in
which the background connection coefficients vanish at that point, so
that $\partial_a = \nabla_a^{\rm B}$.  The result (\ref{eq:Gammaf1})
for general coordinate systems then follows from general covariance.

Next, insert the expansion (\ref{eq:connectioncurved}) of the
connection coefficients into the formula
\begin{equation}
R^a_{\ bcd} = \partial_c \Gamma^a_{bd} - \partial_d \Gamma^a_{bc} +
\Gamma^a_{ce} \Gamma^e_{bd} - \Gamma^a_{de} \Gamma^e_{bc}
\end{equation}
for the Riemann tensor.  Evaluating the result in a coordinate system
in which $\Gamma^{{\rm B}\,a}_{\ \ bc}=0$ at the point of evaluation
gives
\begin{eqnarray}
R^a_{\ bcd} &=& \partial_c \Gamma^{{\rm B}\,a}_{\ \ bd} - \partial_d
\Gamma^{{\rm B}\,a}_{\ \ bc}
+ \varepsilon \left( \partial_c \delta \Gamma^a_{bd} - \partial_d
\delta \Gamma^a_{bc} \right) + O(\varepsilon^2)
\nonumber \\
&=& R^{{\rm B}\,a}_{\ \ \ bcd} + \varepsilon \delta R^a_{\ bcd} +
O(\varepsilon^2)\;.
\label{eq:riemannsplit}
\end{eqnarray}
Here $R^{{\rm B}\,a}_{\ \ \ bcd}$ is the Riemann tensor of the
background metric, and $\delta R^a_{\ bcd} = \partial_c \delta
\Gamma^a_{bd} - \partial_d \delta \Gamma^a_{bc}$ is the linear
perturbation to the Riemann tensor.  It follows from general
covariance that the expression for $\delta R^a_{\ bcd}$ in a general
coordinate system is
\begin{equation}
\delta R^a_{\ bcd} =
\nabla^{{\rm B}}_c \delta \Gamma^a_{bd} - \nabla^{{\rm B}}_d
\delta \Gamma^a_{bc}\;.
\end{equation}
Using the expression (\ref{eq:Gammaf1}) now gives
\begin{eqnarray}
\delta R^a_{\ bcd} &=&
\frac{1}{2} \bigg(
\nabla^{{\rm B}}_c \nabla^{{\rm B}}_b h^a_{\ d}
+ \nabla^{{\rm B}}_c \nabla^{{\rm B}}_d h^a_{\ b}
- \nabla^{{\rm B}}_c \nabla^{{\rm B}\,a} h_{bd}
\nonumber \\ \mbox{} &&
-\nabla^{{\rm B}}_d \nabla^{{\rm B}}_b h^a_{\ c}
- \nabla^{{\rm B}}_d \nabla^{{\rm B}}_c h^a_{\ b}
+ \nabla^{{\rm B}}_d \nabla^{{\rm B}\,a} h_{bc} \bigg)\;.
\end{eqnarray}
Contracting on the indices $a$ and $c$ yields the linearized
Ricci tensor $\delta R_{bd}$:
\begin{equation}
\delta R_{bd} =
- \frac{1}{2}\Box^{\rm B} h_{bd}
-\frac{1}{2} \nabla^{{\rm B}}_d \nabla^{{\rm B}}_b h
+ \nabla^{{\rm B}}_a \nabla^{{\rm B}}_{(b} h^a_{\ d)}\;.
\end{equation}
where $\Box^{\rm B} \equiv \nabla^{{\rm B}}_a \nabla^{{\rm B}\,a}$,
indices are raised and lowered with the background metric, and $h =
h^a_{\ a}$.  Reversing the trace to obtain the linearized Einstein
tensor $\delta G_{bd}$, and writing the result in terms of the
trace-reversed metric perturbation
\begin{equation}
{\bar h}_{ab} = h_{ab} - \frac{1}{2} g^{\rm B}_{ab} g^{{\rm B}\,cd}
h_{cd}
\end{equation}
yields the linearized vacuum Einstein equation
\begin{eqnarray}
0 = \delta G_{bd} &=& - \frac{1}{2} \Box^{\rm B} {\bar h}_{bd} +
R^{\rm B}_{adbc} {\bar h}^{ac}
\nonumber \\ \mbox{} &&
- \frac{1}{2} g^{\rm B}_{bd} \nabla^{\rm B}_a \nabla^{\rm
  B}_c {\bar h}^{ac}
+ \frac{1}{2} \nabla^{\rm B}_b \nabla^{\rm B}_a {\bar h}^a_{\ d}
+ \frac{1}{2} \nabla^{\rm B}_d \nabla^{\rm B}_a {\bar h}^a_{\ b}\;.
\label{eq:lec}
\end{eqnarray}

As in Sec.\ \ref{sec:basic_lin}, the linearized Einstein equation can
be simplified considerably by a suitable choice of gauge.  Under a
gauge transformation parameterized by the vector field $\xi^a$, the
metric transforms as
\begin{equation}
h_{ab} \to h'_{ab} = h_{ab} - 2 \nabla^{\rm B}_{(a} \xi_{b)}\;;
\end{equation}
the divergence of the trace-reversed metric perturbation thus transforms as
\begin{equation}
\nabla^{{\rm B}\,a} {\bar h}'_{ab} = \nabla^{{\rm B}\,a} {\bar h}_{ab}
- \Box^{\rm B} \xi_b\;.
\end{equation}
We can enforce in the new
gauge the transverse condition
\begin{equation}
\nabla^{{\rm B}\,a} {\bar h}'_{ab} =0
\label{eq:transversec}
\end{equation}
by requiring that $\xi_b$ satisfy the wave equation $\Box^{\rm B}
\xi_b=\nabla^{{\rm B}\,a} {\bar h}_{ab}$.  We can further specialize
the gauge to satisfy $h'=0$.  Dropping the primes, the metric
perturbation is thus traceless and transverse:
\begin{equation}
\nabla^{{\rm B}\,a} h_{ab} = h = 0.
\label{eq:ttcurved}
\end{equation}
In this gauge the linearized Einstein equation (\ref{eq:lec}) simplifies to
\begin{equation}
0 = \delta G_{bd} = - \frac{1}{2} \Box^{\rm B} h_{bd} +
R^{\rm B}_{adbc} h^{ac}\;.
\label{eq:lec2}
\end{equation}
(Note, however, that one cannot in this context impose the additional
gauge conditions $h_{0a} = 0$ used in the definition of TT gauge for
perturbations of flat spacetime.)

To see that the traceless condition $h=0$ can be achieved, note that
the trace transforms as
\begin{equation}
h \to h' = h - 2 \nabla^{{\rm B}\,a} \xi_a\;.
\end{equation}
Therefore it is sufficient to find a vector field $\xi^a$ that
satisfies $\Box^{\rm B} \xi_a =0$ and
\begin{equation}
\nabla^{B\,a} \xi_a - h/2=0.
\label{eq:tfcondt}
\end{equation}
We can choose initial data for $\xi_a$ on any Cauchy hypersurface for
which the quantity (\ref{eq:tfcondt}) and also its normal derivative
vanish.  Since the quantity (\ref{eq:tfcondt}) satisfies the
homogeneous wave equation by Eqs.\ (\ref{eq:lec}) and
(\ref{eq:transversec}), it will vanish everywhere.

The wave equation (\ref{eq:lec2}) differs from its flat spacetime
counterpart (\ref{eq:elin1}) in two respects: First, there is an
explicit coupling to the background Riemann tensor; and second, there
is a coupling to the background curvature through the connection
coefficients that appear in the covariant wave operator $\Box^{\rm
B}$.  In the limit (discussed below) where the wavelength of the waves
is much smaller than the lengthscales characterizing the background
metric, these couplings have the effect of causing gradual evolution
in the properties of the wave.  These gradual changes can be described
using the formalism of geometric optics, which shows that GWs travel
along null geodesics with slowly evolving amplitudes and
polarizations.  See Ref.\ \cite{thorne1983} for a detailed description
of this formalism.  Outside of the geometric optics limit the
curvature couplings in Eq.\ (\ref{eq:lec2}) can cause the dynamics of
the metric perturbation to be strongly coupled to the dynamics of the
background spacetime.  An example of such coupling is the parametric
amplification of metric perturbations during inflation in the early
Universe \cite{relicgs}.

\subsection{General definition of gravitational waves: The geometric
  optics regime}

The linear perturbation formalism described in the last section can be
applied to any perturbations of any vacuum background spacetime.  Its
starting point is the separation of the spacetime metric into a
background piece plus a perturbation.  In most circumstances this
separation is merely a mathematical device and can be chosen
arbitrarily; no unique separation is determined by local physical
measurements.  [Although $g_{ab}^{\rm B}$ and $h_{ab}$ are uniquely
determined once one specifies the one parameter family of metrics
$g_{ab}(\varepsilon)$, a given physical situation will be described by
a single metric $g_{ab}(\varepsilon_0)$ for some fixed value of
$\varepsilon_0$ of $\varepsilon$, not by the one parameter family of
metrics.]  However, in special circumstances, a unique separation into
background plus perturbation {\it is} determined by local physical
measurements, and it is only in this context that GWs can be defined.
Such circumstances arise when the wavelength $\lambda$ of the waves is
very much smaller than the characteristic lengthscales ${\cal L}$
characterizing the background curvature.  In this case one can define
the background metric and perturbation, to linear order, via
\begin{eqnarray}
\label{eq:backgrounddef}
g_{ab}^{\rm B} &\equiv& \langle g_{ab} \rangle\;,  \\
\varepsilon h_{ab} &\equiv& g_{ab} - g_{ab}^{\rm B}\;.
\label{eq:pertdef}
\end{eqnarray}
Here the angular brackets $\langle ... \rangle$ denote an average over
lengthscales large compared to $\lambda$ but small compared to ${\cal
L}$; a suitable covariant definition of such averaging has been given
by Brill and Hartle {\cite{bh64}}.  A useful analogy to consider is
the surface of an orange, which contains curvatures on two different
lengthscales: An overall, roughly spherical background curvature
(analogous to the background metric), and a dimpled texture on small
scales (analogous to the GW).  The regime $\lambda \ll {\cal L}$ is
called the {\it geometric optics regime}.

We will argue below that the short-wavelength perturbation
$\varepsilon h_{ab}$ gives rise to an effective stress tensor of order
$\varepsilon^2 h^2 / \lambda^2$, where $h$ is a typical size of
$h_{ab}$.  This effective stress tensor contributes to the curvature
of the background metric $g_{ab}^{\rm B}$.  This contribution to the
curvature is $\lesssim 1/{\cal L}^2$.  It follows that
$\varepsilon^2 h^2 / \lambda^2 \lesssim 1 / {\cal L}^2$, or
\begin{equation}
\varepsilon h \lesssim {\lambda \over {\cal L}}\;.
\end{equation}
Since we are assuming that $\lambda \ll {\cal L}$, it follows that the
short-wavelength piece $\varepsilon h_{ab}$ of the metric is small
compared to the background metric, and so we can use the perturbation
formalism of Sec.\ \ref{sec:pertcurved}.  Consider now the splitting
of the Riemann tensor into a background piece plus a perturbation
given by Eq.\ (\ref{eq:riemannsplit}):
\begin{equation}
R_{abcd} = R^{{\rm B}}_{abcd} + \varepsilon \delta R_{abcd} +
O(\varepsilon^2)\;.
\end{equation}
By the definition (\ref{eq:backgrounddef}) of the background metric,
it follows that $g_{ab}^{\rm B}$ and $R^{{\rm B}}_{abcd}$ vary only
over lengthscales $\gtrsim {\cal L}$, and therefore it follows that to a
good approximation
\begin{equation}
\langle R^{\rm B}_{abcd} \rangle = R^{\rm B}_{abcd}\;.
\end{equation}
Hence the perturbation to the Riemann tensor can be obtained via
\begin{equation}
\varepsilon \delta R_{abcd} = R_{abcd} - \langle R_{abcd} \rangle\;,
\end{equation}
the same unique and local procedure as for the metric perturbation
(\ref{eq:pertdef}).  This Riemann tensor perturbation is often called
the {\it GW Riemann tensor}; it is a tensor characterizing the GWs
that propagates in the background metric $g_{ab}^{\rm B}$.

The operational meaning of the GW fields $\varepsilon h_{ab}$ and
$\varepsilon \delta R_{abcd}$ follows directly from the equivalence
principle and from their meaning in the context of flat spacetime
(Sec.\ \ref{sec:basic_lin}).  Specifically, suppose that ${\cal P}$ is
a point in spacetime, and pick a coordinate system in which
$g_{ab}^{\rm B} = \eta_{ab}$ and $\Gamma^{{\rm B}\,a}_{\ \ bc}=0$ at
${\cal P}$.  Then we have
\begin{equation}
g_{ab} = \eta_{ab} + O \left( {x^2 \over {\cal L}^2 } \right) +
\varepsilon h_{ab} + O(\varepsilon^2)\;,
\end{equation}
where $x$ is the distance from ${\cal P}$.  Therefore, within a
spacetime region around ${\cal P}$ in which $x \ll {\cal L}$, the
flat-spacetime perturbation theory and measurement analysis of Sec.\
{\ref{sec:basic_lin}} can be applied.  Thus, the gravitational
waveforms seen by observers performing local experiments will just be
given by components of the GW Riemann tensor in the observer's local
proper reference frames.

We remark that the splitting of the metric into a background plus a
linear perturbation can sometimes be uniquely defined even in the
regime $\lambda \sim {\cal L}$.  Some examples are when the background
spacetime is static (eg perturbations of a static star), or
homogeneous (eg Friedman-Robertson-Walker cosmological models).  In
these cases the dynamic metric perturbation are not actually GWs,
although their evolution can be computed using the linearized Einstein
equation.  For example, consider the evolution of a metric
perturbation mode which is parametrically amplified during inflation
in the early Universe.  At early times during inflation, the mode's
wavelength $\lambda$ is smaller than the Hubble scale (${\cal L}$);
the mode is said to be ``inside the horizon''.  Any excitation of the
mode is locally measurable (although such modes are usually assumed to
start in their vacuum state).  As inflation proceeds, the mode's
wavelength redshifts and becomes larger due to the the rapid expansion
of the Universe, and eventually becomes larger than the Hubble scale
${\cal L}$; the mode is then ``outside the horizon''.  At this point,
excitations in the mode are not locally measurable and are thus not
GWs.  Finally, after inflation ends the mode ``re-enters the horizon''
and excitations of the mode are locally measurable.  The mode is now a
true GW once again.

Finally, we note that for perturbations of flat spacetime, the
definition of GWs given here does not always coincide with the
definition in terms of the TT component of the metric given in Sec.\
\ref{subsec:gauge_invar}.  However, far from sources of GWs (the
regime relevant to observations), the two definitions do coincide.
This is because the TT piece of the metric will vary on scales of a
wavelength $\lambda$ which is short compared to the lengthscale $\sim
r$ over which other pieces of the metric vary (except for other
dynamic pieces of the metric such as the time-varying quadrupole term
in the gauge-invariant field $\Phi$; those pieces vary on short
lengthscales but are unimportant since they are smaller than the TT
piece by a factor $\sim \lambda^2 / r^2$ or smaller).

\subsection{Effective stress-energy tensor of gravitational waves}
\label{sec:Tab}

Two major conceptual building blocks are needed for the derivation of
the energy and momentum carried by GWs \cite{isaacson}: The
perturbation theory of Sec.\ \ref{sec:pertcurved}, generalized to
second order in $\varepsilon$, and the separation of lengthscales
$\lambda \ll {\cal L}$ discussed in the previous subsection.

We start by discussing the second order perturbation theory.  By
inserting the expansion (\ref{eq:metricexpand}) into the vacuum
Einstein equation we obtain
\begin{eqnarray}
0 &=& G_{ab}
\nonumber \\
&=& G_{ab}[g_{cd}^{\rm B}] + \varepsilon G_{ab}^{(1)}[h_{cd};
  g_{ef}^{\rm B}] + \varepsilon^2 G_{ab}^{(1)}[j_{cd}; g_{ef}^{\rm B}]
  + \varepsilon^2 G_{ab}^{(2)}[h_{cd}; g_{ef}^{\rm B}]
\nonumber\\
& & + O(\varepsilon^3)\;.
\end{eqnarray}
Here $G_{ab}[g_{cd}^{\rm B}]$ is the Einstein tensor of the background
metric, and $G_{ab}^{(1)}[...; g_{ef}^{\rm B}]$ is the linear
differential operator on metric perturbations giving the linear
perturbation to the Einstein tensor generated by a metric
perturbation.  The explicit expression for
$G_{ab}^{(1)}[h_{cd},g_{ef}^{\rm B}]$ is given by Eq.\ (\ref{eq:lec}).
The term $G_{ab}^{(2)}[h_{cd}; g_{ef}^{\rm B}]$ is the piece of the
Einstein tensor that is quadratic in $h_{ab}$; it can be computed by
extending the computation of Sec.\ (\ref{sec:pertcurved}) to one
higher order, and is a sum of terms of the form $h_{ab} \nabla^{\rm
B}_c \nabla^{\rm B}_d h_{ef}$ and $(\nabla_a^{\rm B} h_{bc})
(\nabla^{\rm B}_d h_{ef})$ with various index contractions; see Eq.\
(35.58b) of MTW \cite{mtw}.  It's worth recalling that $j_{ab}$ is a
second order metric perturbation.  We must take the calculation to
second order to compute the effective stress energy tensor of the
waves since an averaging is involved --- the first order contribution
vanishes by the oscillatory nature of the waves.

Equating to zero the coefficients of the different powers of
$\varepsilon$ we obtain the vacuum Einstein equation for the
background spacetime
\begin{equation}
G_{ab}[g^{\rm B}_{cd}] =0,
\label{eq:vacbackground}
\end{equation}
the linearized Einstein equation
\begin{equation}
G_{ab}^{(1)}[h_{cd};
  g_{ef}^{\rm B}] =0\;,
\label{eq:leq3}
\end{equation}
together with the equation for the second-order metric perturbation
$j_{ab}$
\begin{equation}
G_{ab}^{(1)}[j_{cd}; g_{ef}^{\rm B}] =
- G_{ab}^{(2)}[h_{cd}; g_{ef}^{\rm B}]\;.
\label{eq:qeq}
\end{equation}

We now specialize to the geometric optics regime $\lambda\ll{\cal L}$.
We split the second order metric perturbation into a piece $\langle
j_{ab} \rangle$ that is slowly varying, and a piece
\begin{equation}
\Delta j_{ab} = j_{ab} - \langle j_{ab} \rangle
\end{equation}
that is rapidly varying.  The full metric can be now be written
\begin{equation}
g_{ab} = \left( g_{ab}^{\rm B} + \varepsilon^2 \langle j_{ab} \rangle
\right) + \left( \varepsilon h_{ab} + \varepsilon^2 \Delta j_{ab}
\right) + O(\varepsilon^3)\;,
\end{equation}
where the first term varies slowly on lengthscales $\sim {\cal L}$,
and the second term varies rapidly on lengthscales $\sim \lambda$.
Consider next the average of the second-order Einstein equation
(\ref{eq:qeq}).  Using the fact that the averaging operation $\langle
\ldots \rangle$ commutes with derivatives we get
\begin{equation}
G_{ab}^{(1)}[\langle j_{cd} \rangle ; g_{ef}^{\rm B}] =
- \langle G_{ab}^{(2)}[h_{cd}; g_{ef}^{\rm B}] \rangle\;.
\label{eq:qeq1}
\end{equation}
Subtracting Eq.\ (\ref{eq:qeq1}) from Eq.\ (\ref{eq:qeq}) gives an
equation for $\Delta j_{ab}$:
\begin{equation}
G^{(1)}_{ab}[\Delta j_{cd} ] =
- G_{ab}^{(2)}[h_{cd}; g_{ef}^{\rm B}]
+ \langle G_{ab}^{(2)}[h_{cd}; g_{ef}^{\rm B}] \rangle\;.
\end{equation}
Equation (\ref{eq:qeq1}) can be rewritten using Eq.\
(\ref{eq:vacbackground}) as\footnote{Our derivation of the effective
Einstein equation (\protect{\ref{eq:effeq}}) requires the assumption
$\varepsilon^2 \langle j_{ab} \rangle \ll g_{ab}^{\rm B}$, since we
use second order perturbation theory.  However the final result is
valid without this assumption \cite{isaacson}; the curvature generated
by the GWs can be comparable to the background curvature.}
\begin{equation}
G_{ab}[ g_{cd}^{\rm B} + \varepsilon^2 \langle j_{cd} \rangle ] = 8
\pi T_{ab}^{\rm GW,eff} + O(\varepsilon^3)\;,
\label{eq:effeq}
\end{equation}
where the effective GW stress-energy tensor is
\begin{equation}
T_{ab}^{\rm GW,eff} = - {1 \over 8 \pi} \langle G_{ab}^{(2)}[h_{cd};
  g_{ef}^{\rm B}] \rangle\;.
\end{equation}
In the effective Einstein equation (\ref{eq:effeq}), all the
quantities vary slowly, on lengthscales $\sim {\cal L}$.  The
left-hand-side is the Einstein tensor of the slowly varying piece of
the metric.  The right hand side is the effective stress energy
tensor, obtained by taking an average of the quadratic piece of the
second order Einstein tensor.  It follows from Eq.\ (\ref{eq:effeq})
that $T_{ab}^{\rm GW,eff}$ is conserved with respect to the metric
$g_{ab}^{\rm B} + \varepsilon^2 \langle j_{ab} \rangle$.  In
particular, to leading order in $\varepsilon$ it is conserved with
respect to the background metric $g_{ab}^{\rm B}$.

The effect of the GWs is thus to give rise to a correction $\langle
j_{ab} \rangle$ to the background metric.  This correction is locally
of the same order as $\Delta j_{ab}$, the rapidly varying piece of the
second order metric perturbation.  However, any measurements that
probe only the long-lengthscale structure of the metric (for example
measurements of the gravitating mass of a radiating source over
timescales long compared to $\lambda$) are sensitive only to $\langle
j_{ab} \rangle$.  Thus, when one restricts attention to long
lengthscales, GWs can thus be treated as any other form of matter
source in general relativity.  Typically $\langle j_{ab} \rangle$
grows secularly with time, while $\Delta j_{ab}$ does not.

A fairly simple expression for the effective stress-energy tensor can
be obtained as follows.  Schematically, the effective stress-energy
tensor has the form
\begin{equation}
T_{ab}^{\rm GW, eff}\sim \& \langle h_{ab} \nabla^{\rm B}_c
\nabla^{\rm B}_d h_{ef} \rangle + \& \langle (\nabla_a^{\rm B} h_{bc})
(\nabla^{\rm B}_d h_{ef} )\rangle\;,
\label{eq:schematic}
\end{equation}
where $\&$ means ``a sum of terms obtained by taking various
contractions of''.  In this expression gradients scale as $1/\lambda$,
so $\nabla_c^{\rm B} \nabla_d^{\rm B} \sim 1/\lambda^2$.  However the
commutator of two derivatives scales as the background Riemann tensor,
which is of order $1/{\cal L}^2$.  Therefore, up to corrections of
order $\lambda^2/{\cal L}^2$ which can be neglected, one can freely
commute covariant derivatives in the expression (\ref{eq:schematic}).
Also, the average of any total derivative will vanish in the limit
$\lambda \ll {\cal L}$ if the averaging lengthscale is taken to be
$\sqrt{\lambda {\cal L}}$.  Therefore one can flip derivatives from
one factor to another inside the averages in Eq.\
(\ref{eq:schematic}), as in integration by parts.  Using these
manipulations the expression for the effective stress-energy tensor
simplifies to {\cite{mtw,isaacson}}
\begin{eqnarray}
\fl
T_{ab}^{\rm GW,eff} = \frac{1}{32 \pi} \left< \nabla^{\rm B}_a {\bar
  h}_{cd} \nabla_b^{\rm B} {\bar h}^{cd} - \frac{1}{2}
\nabla^{\rm B}_a  {\bar h} \nabla^{\rm B}_b  {\bar h}
- \nabla^{\rm B}_a {\bar h}_{bc} \nabla_d^{\rm B} {\bar h}^{cd}
- \nabla^{\rm B}_b {\bar h}_{ac} \nabla_d^{\rm B} {\bar h}^{cd}
\right>\;.
\end{eqnarray}
In gauges satisfying the transverse-traceless conditions
(\ref{eq:ttcurved}) this reduces to
\begin{eqnarray}
T_{ab}^{\rm GW,eff} &=& \frac{1}{32 \pi} \left< \nabla^{\rm B}_a
h_{cd} \nabla_b^{\rm B} h^{cd} \right>.
\end{eqnarray}
For example, for the plane wave propagating in the $z$ direction in
flat spacetime, given by
\begin{eqnarray}
h_{xx} &=& -h_{yy} = h_0\cos(\omega t - \omega z)\;,
\nonumber\\
h_{ab} &=& 0\quad\mbox{(all other components)}\;,
\end{eqnarray}
the energy density and energy flux are given by
\begin{equation}
T^{tt} = T^{tz} =
{h_0^2 \omega^2 \over 16 \pi} \langle \cos^2(\omega t - \omega z)
\rangle =
{h_0^2 \omega^2 \over 32 \pi}.
\end{equation}
If we restore factors of $G$ and $c$, and insert numbers typical of
bursts of waves that we hope to detect, we get the energy flux
\begin{equation}
T^{tz} = 1.5 \,\, {\rm mW} \,\, {\rm m}^{-2} \left(
\frac{h_0}{10^{-22}} \right)^2 \, \left( \frac{f}{1 \, {\rm kHz}}
\right)^2\;
\end{equation}
where $f = \omega / (2 \pi)$.  Note that this is a large energy flux
by astronomical standards, despite the tiny value of $h_0$; it is
comparable to the flux of reflected sunlight from a full moon
\cite{schutz2001}.

\section{A brief survey of gravitational wave astronomy}
\label{sec:synopsis}

Having now reviewed the basic theory and properties of GWs, we
conclude this article by very briefly surveying the properties of
important potential sources of GWs.  Our goal is to give some
indication of the value that GWs may provide for astronomical
observations; much of this material is updated from a previous survey
article, Ref.\ {\cite{annals}}.  We note that since the focus of this
article is intended to be the theory GW sources (and that this article
is {\it significantly} longer than was intended or requested), we are
quite a bit more schematic in our treatment here than we have been in
the rest of this paper.
%  We apologize to the reader for this change in
%style, but hope it is understood that 
This final section is intended to be a very brief, somewhat
superficial survey, rather than a detailed review. 

We begin by contrasting gravitational radiation with electromagnetic
radiation, which forms the basis for almost all current astronomical
observations:

\medskip
\noindent {\it Electromagnetic waves interact strongly with matter;
GWs do not.}  The weak interaction of GWs is both blessing and curse:
It means that they propagate from emission to Earth-bound observers
with essentially zero absorption, making it possible to probe
astrophysics that is hidden or dark to electromagnetic observations
--- e.g., the coalescence and merger of black holes, the collapse of a
stellar core, the dynamics of the early Universe.  It also means that
detecting GWs is very difficult.  Also, because many of the best
sources are hidden or dark, they are very poorly understood today ---
we know very little about what are likely to be some of the most
important sources of GWs.

\medskip
\noindent {\it Electromagnetic radiation typically has a wavelength
smaller than the size of the emitting system, and so can be used to
form an image of the source.}  This is because electromagnetic
radiation is usually generated by moving charges in the environment of
some larger source --- e.g., an atomic transition in interstellar gas,
or emission from hot plasma in a stellar environment.  {\it By
contrast, the wavelength of gravitational radiation is typically
comparable to or larger than the size of the radiating source.}  GWs
are generated by the bulk dynamics of the source itself --- e.g., the
motion of neutron stars in a binary.  As a consequence, GWs {\it
cannot} be used to form an image: The radiation simply does not
resolve the generating system.  Instead, GWs are best thought of as
analogous to sound --- the two polarizations carry a stereophonic
description of the source's dynamics.

\medskip
\noindent {\it Gravitons in a gravitational-wave burst are phase
coherent; photons in electromagnetic signals are usually
phase-incoherent.}  This arises from the fact that each graviton is
generated from the same bulk motion of matter or of spacetime
curvature, while each photon is normally generated by different,
independent events involving atoms, ions or electrons.  Thus GWs are
similar to laser light.  We can take advantage of the phase coherence
of GWs to enhance their detectability.  Matched filtering techniques
for detecting GW bursts with well-modeled functional form (like those
generated by coalescing compact binaries) extend the distance to which
sources can be seen by a factor of roughly the square root of the
number of cycles in the waveform, a significant gain \cite{300yrs}.

\medskip
\noindent
An extremely important consequence of this coherency is that {\it the
direct observable of gravitational radiation is the strain $h$, a
quantity that falls off with distance as $1/r$.}  Most electromagnetic
observables are some kind of energy flux, and so fall off with a
$1/r^2$ law; measuring coherent GWs is analogous to measuring a
coherent, $1/r$ electromagnetic radiation field.  This comparatively
slow fall off with radius means that relatively small improvements in
the sensitivity of GW detectors can have a large impact on their
science: Doubling the sensitivity of a detector doubles the distance
to which sources can be detected, increasing the volume of the
Universe to which sources are measurable by a factor of 8.  Every
factor of two improvement in the sensitivity of a GW observatory
should increase the number of observable sources by about an order of
magnitude.

\medskip
\noindent {\it In most cases, electromagnetic astronomy is based on
deep imaging of small fields of view: Observers obtain a large amount
of information about sources on a small piece of the sky.  GW
astronomy will be a nearly all-sky affair: GW detectors have nearly
$4\pi$ steradian sensitivity to events over the sky.}  A consequence
of this is that their ability to localize a source on the sky is not
good by usual astronomical standards; but, it means that any source on
the sky will be detectable, not just sources towards which the
detector is ``pointed''.  The contrast between the all-sky sensitivity
but poor angular resolution of GW observatories, and the pointed, high
angular resolution of telescopes is very similar to the angular
resolution contrast of hearing and sight, strengthening the useful
analogy of GWs with sound.

\medskip
From these general considerations, we turn now to specifics.  It is
useful to categorize GW sources (and the methods for detecting their
waves) by the frequency band in which they radiate.  Broadly speaking,
we may break the GW spectrum into four rather different bands: the
{\it ultra low frequency} band, $10^{-18}\,{\rm Hz}\lesssim f\lesssim
10^{-13}\,{\rm Hz}$; the {\it very low frequency} band, $10^{-9}\,{\rm
Hz}\lesssim f\lesssim 10^{-7}\,{\rm Hz}$; the {\it low frequency}
band, $10^{-5}\,{\rm Hz}\lesssim f\lesssim 1\,{\rm Hz}$; and the {\it
high frequency} band, $1\,{\rm Hz}\lesssim f\lesssim 10^4\,{\rm Hz}$.

For compact sources, the GW frequency band is typically related to the
source's size $R$ and mass $M$.  Here the source size $R$ means the
lengthscale over which the source's dynamics vary; for example, it
could be the actual size of a particular body, or the separation of
members of a binary.  The ``natural'' GW frequency of such a source is
$f_{\rm GW} \sim (1/2\pi)\sqrt{G M/R^3}$.  Because $R\gtrsim 2 G
M/c^2$ (the Schwarzschild radius of a mass $M$), we can estimate an
upper bound for the frequency of a compact source:
\begin{equation}
f_{\rm GW}(M) < {1\over4\sqrt{2}\pi}{c^3\over G M} \simeq 10^4\,{\rm
Hz} \left({M_\odot\over M}\right)\;.
\label{eq:fgw_approx}
\end{equation}
This is a rather hard upper limit, since many interesting sources are
quite a bit larger than $2 G M/c^2$, or else evolve through a range of
sizes before terminating their emission at $R \sim 2 GM/c^2$.
Nonetheless, this frequency gives some sense of the types of compact
sources that are likely to be important in each band --- for example,
high frequency compact sources are of stellar mass (several solar
masses); low frequency compact sources are thousands to millions of
solar masses, or else contain widely separated stellar mass bodies.

\subsection{High frequency}
\label{subsec:high}

The high frequency band, $1\,{\rm Hz}\lesssim f \lesssim 10^4\,{\rm
Hz}$, is targeted by the new generation of ground-based laser
interferometric detectors such as LIGO.  (It also corresponds roughly
to the audio band of the human ear: When converted to sound, LIGO
sources are audible to humans.)  The low frequency end of this band is
set by the fact that it is extremely difficult to prevent mechanical
coupling of the detector to ground vibrations at low frequencies, and
probably impossible to prevent gravitational coupling to ground
vibrations, human activity, and atmospheric motions
{\cite{ht1998,tw1999,tcreighton}}.  The high end of the band is set by
the fact that it is unlikely any interesting GW source radiates at
frequencies higher than a few kilohertz.  Such a source would have to
be relatively low mass ($\lesssim 1 M_\odot$) but extremely compact
[cf.\ Eq.\ (\ref{eq:fgw_approx})].  There are no known theoretical or
observational indications that gravitationally collapsed objects in
this mass range exist.

The article by Danzmann in this volume {\cite{kd_here}} discusses the
detectors relevant to this frequency band in some detail; our
discussion here is limited to a brief survey of these instruments.
Several interferometric GW observatories are either operating or being
completed in the United States, Europe, Japan, and Australia.  Having
multiple observatories widely scattered over the globe is extremely
important: The multiplicity gives rise to cross-checks that increase
detection confidence and also aids in the interpretation of
measurements.  For example, sky location determination and concomitant
measurement of the distance to a source follows from triangulation of
time-of-flight differences between separated detectors.

The major interferometer projects are:

\begin{itemize}

\item {\bf LIGO.}  The Laser Interferometer Gravitational-wave
Observatory\cite{ligo} consists of three operating interferometers: A
single four kilometer interferometer in Livingston, Louisiana, as well
as a pair of interferometers (four kilometers and two kilometers) in
the LIGO facility at Hanford, Washington.  The sites are separated by
roughly 3000 kilometers, and are situated to support coincidence
analysis of events.

\item {\bf Virgo.} Virgo is a three kilometer French-Italian detector
under construction near Pisa, Italy {\cite{virgo}}.  In most respects,
Virgo is quite similar to LIGO.  A major difference is that Virgo
employs a very sophisticated seismic isolation system that promises
extremely good low frequency sensitivity.

\item {\bf GEO600.} GEO600 is a six hundred meter interferometer
constructed by a German-English collaboration near Hanover, Germany
{\cite{geo}}.  Despite its shorter arms, GEO600 achieves sensitivity
comparable to the multi-kilometer instruments using advanced
interferometry techniques.  This makes it an invaluable testbed for
technology to be used in later generations of the larger instruments,
as well as enabling it to make astrophysically interesting
measurements.

\item {\bf TAMA300.} TAMA300 is a three hundred meter interferometer
operating near Tokyo.  It has been in operation for several years now
{\cite{tama}}.  The TAMA team is currently designing a three kilometer
interferometer {\cite{advtama}}, building on their experiences with
the three hundred meter instrument.

\item {\bf ACIGA.} The Australian Consortium for Interferometric
Gravitational-Wave Astronomy is currently constructing an eighty meter
research interferometer near Perth, Australia {\cite{aciga}}, hoping
that it will be possible to extend it to multi-kilometer scale in the
future.  Such a detector would likely be a particularly valuable
addition to the worldwide stable of detectors, since all the Northern
Hemisphere detectors lie very nearly on a common plane.  An Australian
detector would be far outside this plane, allowing it to play an
important role in determining the location of sources on the sky.

\end{itemize}

The LIGO, GEO, and TAMA instruments have now been operating for
several years; see Refs.\ {\cite{ligo1,ligo2,ligo3,ligo4}} for the
results and upper limits from the first set of observations.
All of these detectors have or will have sensitivities similar to that
illustrated in Fig.\ {\ref{fig:ligo_sens}} (which shows, in
particular, the sensitivity goal of the first generation of LIGO
interferometers).  This figure also shows the ``facility limits'' ---
the lowest noise levels that can be achieved even in principle within
an interferometer facility.  The low level facility limits come from
{\it gravity-gradient noise}: noise arising from gravitational
coupling to fluctuations in the local mass distribution (such as from
seismic motions in the earth near the test masses {\cite{ht1998}},
human activity near the detector {\cite{tw1999}}, and density
fluctuations in the atmosphere {\cite{tcreighton}}).  At higher
frequencies, the facility limit arises from residual gas (mostly
hydrogen) in the interferometer vacuum system --- stray molecules of
gas effectively cause stochastic fluctuations in the index of
refraction.

\begin{figure}[t]
\includegraphics[width = 13cm]{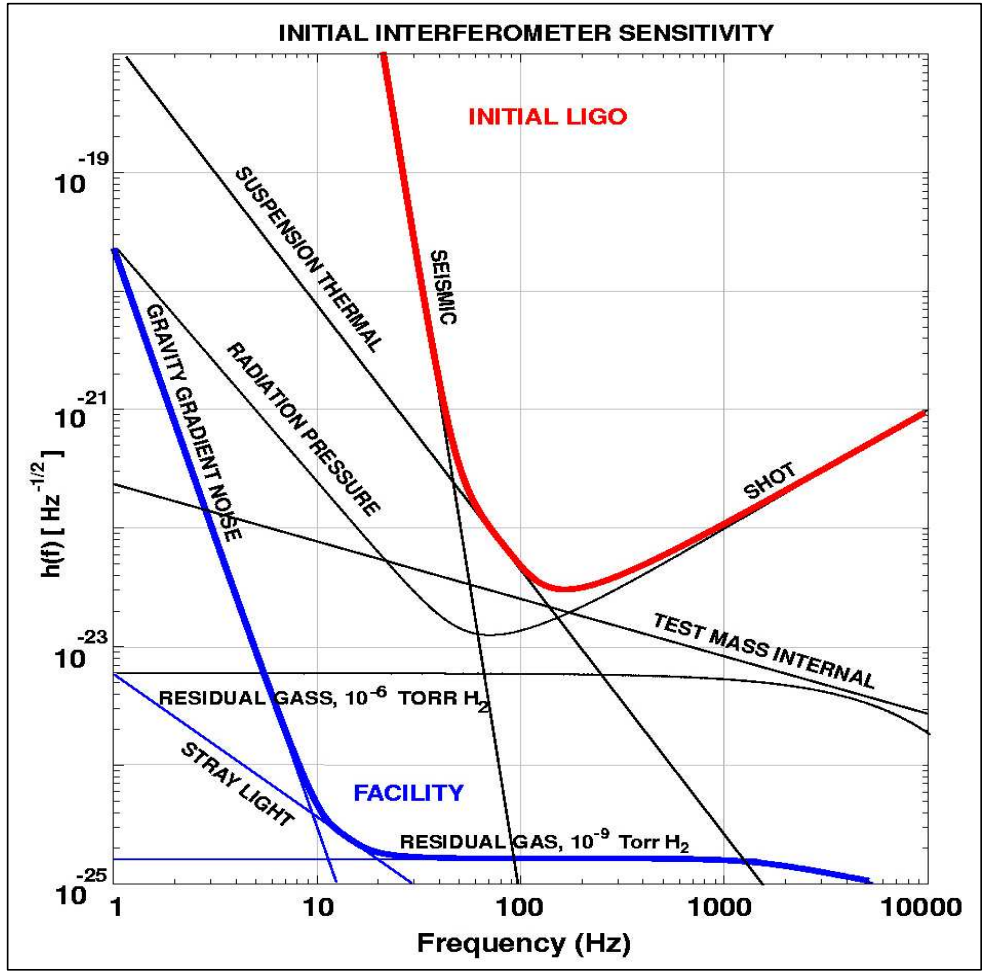}
\caption{Sensitivity goals of the initial LIGO interferometers, and
facility limits on the LIGO sensitivity (taken from Ref.\
{\cite{snowmass}}).}
\label{fig:ligo_sens}
\end{figure}

We now survey the more well-understood possible sources of measurable
GWs in the high-frequency band.  We emphasize at this point that such
a listing of sources can in no way be considered comprehensive: We are
hopeful that some GW sources may surprise us, as has been the case
whenever we have studied the Universe with a new type of radiation.

\subsubsection{Coalescing compact binaries}

Compact binaries --- binary star systems in which each member is a
neutron star or black hole --- are currently the best understood
sources of GWs.  Double neutron stars have been studied
observationally since the mid 1970s; five such systems
{\cite{wt04,sttw02,dk94,betal03,fetal04}} tight enough to merge within
a few $10^8$ or $10^9$ years have been identified in our Galaxy.
Extrapolation from these observed binaries in the Milky Way to the
Universe at large {\cite{nps,phinney91,kl2000,burgay}} indicates that
GW detectors should measure at least several and at most several
hundred binary neutron star mergers each year (following detector
upgrades; the expected rate for initial detectors is of order one
event per several years, so that measurement of an event is plausible
but of fairly low probability).  Population synthesis (modeling
evolution of stellar populations) indicates that the measured rate of
binaries containing black holes should likewise be interestingly large
(perhaps even for initial detectors)
{\cite{bb98,pzy98,pzw2000,bkb2002}}.  The uncertainties of population
synthesis calculations are rather large, however, due to poorly
understood aspects of stellar evolution and compact binary formation;
data from GW detectors is likely to have a large impact on this field.

\subsubsection{Stellar core collapse}

Core collapse in massive stars (the engine of Type II supernova
explosions) has long been regarded as likely to be an important source
of GWs; see, for example, Ref.\ {\cite{eardley83}} for an early
review.  Stellar collapse certainly exhibits all of the {\it
necessary} conditions for strong GW generation --- large amounts of
mass ($1 - 100\,M_\odot$) flow in a compact region (hundreds to
thousands of kilometers) at relativistic speeds ($v/c \gtrsim 1/5$).
However, these conditions are not {\it sufficient} to guarantee strong
emission.  In particular, the degree of asymmetry in collapse is not
particularly well understood.

If the core of a star is very rapidly rotating during collapse, then
instabilities may develop which lead to strong GW emission
{\cite{c69}}.  If such instabilities develop, core collapse GWs could
be detected from events as far away as 10 Megaparsecs {\cite{fhh02}},
a distance encompassing enough galaxies that several events per year
would be likely.  Most models of massive stars, however, indicate that
such rapid rotation is not likely (e.g., {\cite{fh00}}).  Even without
the growth of instabilities, the asymmetric dynamics of core collapse
is likely to lead to wave emission which would be detectable within
the Local Group of galaxies, with perhaps an event every few years
detectable by advanced interferometers {\cite{dfm02}}.  The wave
strength is likely to correlate strongly with the degree of asymmetry in
the supernova.  If the GW event has an
electromagnetic or neutrino counterpart we may gain a wealth of
knowledge regarding the state of the precollapse core {\cite{fhh04}}.

\subsubsection{Periodic emitters}

Periodic sources of GWs radiate at constant or nearly constant
frequency, like radio pulsars.  In fact, the prototypical source of
continuous GW is a rotating neutron star, or GW pulsar.  A
non-axisymmetry in a neutron star crust (caused, for example, by an
oblateness that is misaligned with the star's spin axis) will radiate
GWs with characteristic amplitude
\begin{equation}
h \sim {G\over c^4}{I f^2 \epsilon \over r}\;.
\end{equation}
Here $I$ is the star's moment of inertia, $f$ is the wave frequency,
$r$ is the distance to the source, and $\epsilon$ is the dimensionless
fractional distortion $\epsilon = (I_{xx} - I_{yy}) / I$, where
$I_{ij}$ is the moment of inertia tensor.  The crucial parameter
$\epsilon$ characterizes the degree to which the star is distorted; it
is rather poorly understood.  Upgraded interferometers in LIGO could
set an upper limit on $\epsilon$ of order $10^{-6}$ for sources at
$\sim 10$ kpc \cite{ct02}.  Various mechanisms have been proposed to
explain how a neutron star can be distorted to give a value of
$\epsilon$ that is interesting as a GW source; see {\cite{jones,ben}}
for further discussion.  Examples of some interesting mechanisms
include misalignment of a star's internal magnetic field with the
rotation axis {\cite{cutler02}} and distortion by accreting material
from a companion star {\cite{lars,ucb}} (discussed in more detail
below).

Whatever the mechanism generating the distortion, it is clear that
$\epsilon$ will be small, so that $h \sim 10^{-24}$ or smaller ---
quite weak.  Measuring these waves will require coherently tracking
their signal for a large number of wave cycles.  Coherently tracking
$N$ cycles boosts the signal to noise ratio by a factor
$\sim\sqrt{N}$.  This is actually fairly difficult, since the signal
is strongly modulated by the Earth's rotation and orbital motion,
``smearing'' the waves' power across multiple frequency bands.
Searching for periodic GWs means demodulating the motion of the
detector, a computationally intensive problem since the modulation is
different for every sky position.  Unless one knows in advance the
position of the source, one needs to search over a huge number of sky
position ``error boxes'', perhaps as many as $10^{13}$.  One rapidly
becomes computationally limited\footnote{This rather large number of
patches on the sky is driven by the possible need to search for high
frequency pulsars over several months of observation.  The difference
$\Delta f$ between the Doppler frequency shifts for two adjacant sky
patches separated by an angle $\delta \theta$ is of order $\Delta
f\sim v_\oplus f \delta \theta /c$, where $v_\oplus \sim 3 \times 10^4
\, {\rm m} {\rm s}^{-1}$ is the Earth's orbital velocity and $f$ is
the gravitational wave frequency.  The phase error over an observation
time $T_{\rm obs}$ is of order $\Delta f T_{\rm obs}$.  Demanding that
this be less than unity yields $\delta \theta \lesssim c / (v_\oplus f
T)$.  The number of independent sky patches is then $N_{\rm p} \sim 4
\pi \delta \theta^{-2} \sim 4 \pi v_\oplus^2 f^2 T_{\rm obs}^2 / c^2
\sim 10^{13}$ for $f = 1000 \, {\rm Hz}$ and $T_{\rm obs} = 1/3$ year.
Fewer positions would be needed if either the maximum frequency or the
integration time is reduced; the figures given here set the maximum
values that are plausible.  See Ref. \cite{periodic} for more
details.}.  (Radio pulsar searches face this same problem, with the
additional complication that radio pulses are dispersed by the
interstellar medium.  However, radio observers usually use shorter
integration times, and often target their searches on small regions of
the sky, so their computational cost is usually not as great.)  For
further discussion, see {\cite{periodic}}; for ideas about doing
hierarchical searches that require less computer power, see
{\cite{per_hier}}.

As mentioned above, one particularly interesting mechanism for
distorting a neutron star comes from accretion of material from a
companion star.  Accretion provides a spin-up torque to a neutron
star,
\begin{equation}
\left(dJ/dt\right)_{\rm spin-up} \sim R^2\Omega_* \dot M
\label{eq:torque1}
\end{equation}
(where $J$ is the spin angular momentum, $\Omega_*$ is the orbital
frequency of the accreting matter as it plunges onto the star, $R$ is
the star's radius, and $\dot M$ is the mass accretion rate).  Without
any kind of braking mechanism, the neutron star would presumably
spin-up until it reaches the ``breakup limit'', i.e., the spin
frequency at which centrifugal forces would begin to break it apart;
the breakup frequency is typically around 2000 -- 3000 Hz.

Observations have shown {\cite{d03}} that accreting neutron stars do,
in fact, appear to have a ``speed limit'' --- no accreting neutron
star has been observed to spin faster than 619 Hz {\cite{d04}}.  This
is consistent with the fact that the fastest {\it radio}
pulsar\footnote{The so-called ``recycled'' radio pulsars spin at
frequencies $\sim$ several hundred Hertz; they are believed to be the
fossils of accreting neutron stars.} has a spin period of 641 Hz
{\cite{betal82}}.  This suggest that {\it some} mechanism is removing
angular momentum from the neutron star.  A plausible and very
attractive possibility for how this angular momentum is removed is via
GW emission.  Because the spin-down torque due to GW emission grows
sharply with spin frequency,
\begin{equation}
\left(dJ/dt\right)_{\rm spin-down} \propto \Omega^5
\quad\mbox{(Quadrupole emission)}\;,
\label{eq:torque2}
\end{equation}
the limiting spin obtained by balancing the torques (\ref{eq:torque1})
and (\ref{eq:torque2}) is relatively insensitive to the mass accretion
rate ${\dot M}$.  Such a mechanism was originally suggested by Wagoner
{\cite{w84}}, and was revived by Bildsten {\cite{lars}} to explain the
narrow clustering in spin frequency of accreting low-mass x-ray
binaries (LMXBs).  Various mechanisms could provide the spin-down
torque --- Bildsten originally suggested a quadrupole moment in the
spinning star could be induced by a thermally varying electron capture
mechanism, but also noted that the r-mode instability (see, e.g.,
{\cite{ak01}} for a review) could be excited, leading to a similar
spin-down law.  Whatever the mechanism, accreting neutron stars are
obvious and very attractive targets for observing campaigns with GW
detectors, particularly given that their sky positions are well known.

\subsubsection{Stochastic backgrounds}
\label{sec:stochastic}

Stochastic backgrounds are ``random'' GWs, arising from a large number
of independent, uncorrelated sources that are not individually
resolvable.  A particularly interesting source of stochastic waves is
the dynamics of the early Universe, which could produce an all-sky GW
background, similar to the cosmic microwave background; see Refs.\
\cite{bruce96,maggiore,ab03} for detailed reviews.  Stochastic waves
can be generated in the early Universe via a variety of mechanisms:
amplification of primordial fluctuations in the Universe's geometry
via inflation, phase transitions as previously unified interactions
separate, a network of vibrating cosmic strings, or the condensation
of a brane from a higher dimensional space.  These waves can actually
extend over a wide range of frequency bands; waves from inflation in
particular span all bands, from ultra low frequency to high frequency.

Stochastic backgrounds are usually idealized as being stationary,
isotropic and homogeneous.  They are thus characterized by their
energy density per unit frequency, $d \rho_{\rm gw} / df$.  This is
often parameterized in terms of the energy density per unit
logarithmic frequency divided by the critical energy density to close
the Universe
\begin{equation}
\Omega_{\rm gw}(f) = {1\over\rho_{\rm crit}}{d\rho_{\rm gw}\over
d\ln f}\;,
\label{eq:stoch_gw_om}
\end{equation}
where $\rho_{\rm crit} = 3 H_0^2/8\pi G$ is the critical density and
$H_0$ is the value of the Hubble constant today.  Different
cosmological sources produce different levels of $\Omega_{\rm gw}(f)$,
centered in different bands.  In the high frequency band, waves
produced by inflation are likely to be rather weak: estimates suggest
that the spectrum will be flat across LIGO's band, with magnitude
$\Omega_{\rm gw} \sim 10^{-15}$ at best {\cite{turner1}}.  Waves from
phase transitions can be significantly stronger, but are typically
peaked around a frequency that depends on the temperature $T$ of the
phase transition {\cite{bruce96,kmk2001}}:
\begin{equation}
f_{\rm peak} \sim 100\,{\rm Hz}\left({T\over10^5\,{\rm TeV}}\right)\;.
\label{eq:freq_phasetrans}
\end{equation}

Because of their random nature, stochastic GWs look just like noise.
Ground-based detectors will measure stochastic backgrounds by
comparing data at multiple sites and looking for ``noise'' that is
correlated {\cite{maggiore,ar}}.  For comparing to a detector's noise,
one should construct the characteristic stochastic wave strain,
\begin{equation}
h \propto f^{-3/2} \sqrt{\Omega_{\rm gw}(f) \Delta f}\;,
\end{equation}
where $\Delta f$ is the frequency band across which the measurement is
made.  For further discussion and the proportionality constants, see
{\cite{maggiore}}.  Note that if $\Omega_{\rm gw}(f)$ is constant,
this strain level grows sharply with decreasing frequency --- the most
interesting limits are likely to be set by measurements at low
frequencies.

Early detectors will have fairly poor sensitivity to the background,
constraining it to a level $\Omega_{\rm gw}\sim 5\times10^{-6}$ in a
band from about 100 Hz to 1000 Hz.  This is barely more sensitive than
known limits from cosmic nucleosynthesis {\cite{bruce96}}.  Later
upgrades will be significantly more sensitive, able to detect waves
with $\Omega_{\rm gw}\sim 10^{-10}$, which is good enough to place
interesting limits on backgrounds from some phase transitions.

\subsection{Low frequency}
\label{subsec:low}

There is no hope of measuring GWs in the low frequency band,
$10^{-5}\,{\rm Hz}\lesssim f\lesssim 1\,{\rm Hz}$, using a
ground-based instrument: Even if it were possible to completely
isolate one's instrument from local ground motions, gravitational
coupling to fluctuations in the local mass distribution ultimately
limits the sensitivity to frequencies $f \gtrsim 1\,{\rm Hz}$.
Nonetheless, many extremely interesting sources radiate in this band.
The only way to measure these waves is to build a detector in the
quiet environment of space, far removed from low-frequency noise
sources.

Such an instrument is currently being designed jointly by NASA in the
United States and ESA, the European Space Agency: LISA, the Laser
Interferometer Space Antenna.  If all goes well, LISA will be
launched into orbit in 2013 or so.  LISA will be a laser
interferometer, similar in concept to the ground-based detectors:
Changes in the distance between widely separated test masses will be
monitored for GWs.  However, LISA's scale is {\it vastly} larger than
that of its ground-based cousins, and so details of its
operations are quite different.  In particular, LISA will have
armlengths $L \simeq 5 \times 10^6\,{\rm km}$.  The three spacecraft
which delineate the ends of LISA's arms are placed into orbits such
that LISA forms a triangular constellation orbiting the sun, inclined
$60^\circ$ with respect to the plane of the ecliptic and following the
Earth with a $20^\circ$ lag.  This configuration is sketched in Fig.\
{\ref{fig:lisa_orbit}}.  Since it essentially shares Earth's orbit,
the constellation orbits the sun once per year, ``rolling'' as it does
so.  This motion plays an important role in pinpointing the position
of sources by modulating the measured waveform --- the modulation
encodes source location and makes position determination possible.

\begin{figure}[t]
\includegraphics[width = 13cm]{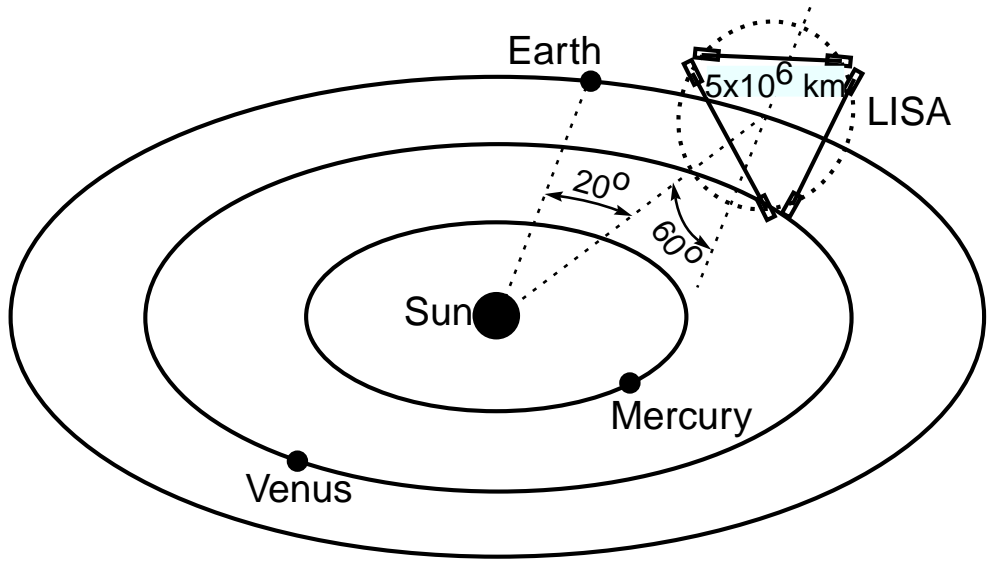}
\caption{Orbital configuration of the LISA antenna.}
\label{fig:lisa_orbit}
\end{figure}

The 3 spacecraft each contain two optical assemblies, each of which
houses a 1 Watt laser and a 30 centimeter telescope.  Because of the
extreme lengths of the interferometer's arms, Fabry-Perot
interferometry as in the ground-based detectors is not possible:
Diffraction spreads the laser beam over a diameter of about $20\,{\rm
km}$ as it propagates from one spacecraft to the other.  A portion of
that $20\,{\rm km}$ wavefront is sampled with the telescope.  That
light is then interfered with a sample of light from the on-board
laser.  Each spacecraft thus generates two interference data streams;
six signals are generated by the full LISA constellation.  From these
six signals, we can construct the time variations of LISA's armlengths
and then build both GW polarizations.  More information and details
can be found in {\cite{tdi}}.

Note that the LISA armlengths are {\it not} constant --- as the
constellation orbits, the distances between the various spacecraft
vary by about $1\%$ (including effects such as planetary
perturbations).  These variations are {\it far} larger than the
displacements produced by GWs, which are of order picometers.
However, these variations occur over timescales of order months, and
are extremely smooth and well modeled.  It will not be difficult to
remove them from the data leaving clean data in the interesting
frequency band.  Picometer scale variations are not too difficult to
measure in this band by gathering photons for a time $10\,{\rm sec}
\lesssim T \lesssim 1\,{\rm day}$.  Even though the bulk of the
laser's emitted power is lost by diffraction, enough photons are
gathered on this timescale that the phase shift due to the GW can be
determined.

The GW signals are actually read out by monitoring the position of the
so-called ``gravitational sensor'' on each optical assembly; in
particular, the position of a ``proof mass'' which floats freely and
constitutes the test mass for the LISA antenna is monitored.  Because
it is freely floating, the proof mass follows a geodesic of the
spacetime.  Micronewton thrusters keep the bulk spacecraft centered on
these proof masses, forcing the craft to follow their average
trajectory.  In this way, the spacecraft are isolated from low
frequency forces that could impact the ability to measure GWs (e.g.,
variations in solar radiation pressure).

We now take a quick tour through some interesting LISA sources:

\subsubsection{Periodic emitters}

In the high-frequency band, the source of most periodic GWs is
expected to be isolated neutron stars.  LISA's periodic GWs will come
primarily from binary star systems in the Milky Way, primarily close
white dwarf binaries.  Most of these systems do not generate waves
strong enough to backreact significantly, so that their frequencies do
not change measurably over the course of LISA observations.  Certain
systems are well-known in advance to be sources of periodic waves for
the LISA band.  These sources are understood well enough from optical
observations that they may be regarded as ``calibrators'' --- LISA had
better detect them, or else something is wrong!

Aside from these sources that are known in advance, it is expected
that LISA will discover a good number of binary systems that are too
faint to detect with telescopes.  Joint observations by LISA and other
astronomical instruments are likely to be more fruitful than
observations with a single instrument alone.  For example, it is
typically difficult for telescopes to determine the inclination of a
binary to the line of sight (a quantity needed to help pin down the
masses of the binary's members).  GWs measure the inclination angle
almost automatically, since this angle determines the relative
magnitude of the polarizations $h_+$ and $h_\times$.

The total number of periodic binaries radiating in LISA's band is
expected to be so large that they will constitute a confused,
stochastic background at low frequencies --- there are likely to be
several thousand galactic binaries radiating in each resolvable
frequency bin.  This background will constitute a source of ``noise''
(from the standpoint of measuring other astrophysical sources) that is
largely than that intrinsic to the instrument noise at $f \lesssim
10^{-3}$ Hz.

\subsubsection{Coalescing binary systems containing black holes}

Coalescing binary black hole systems will be measurable by LISA to
extremely large distances --- essentially to the edge of the
observable Universe.  Even if such events are very rare, the observed
volume is enormous so an interesting event detection rate is very
likely.  One class of such binaries consists of systems in which the
member holes are of roughly equal mass ($\sim 10^5 - 10^8 M_\odot$).
These binaries can form following the merger of galaxies (or
pregalactic structures) containing a black holes in their cores.
Depending on the mass of the binary, the waves from these coalescences
will be detectable to fairly large redshifts ($z \sim 5 - 10$),
possibly probing an early epoch in the formation of the Universe's
structure {\cite{hmnras}}.

The other major class of binary systems consists of relatively small
bodies (black holes with mass $\sim 10$ -- $100\,M_\odot$, neutron
stars, or white dwarfs) that are captured by larger black holes
($M\sim 10^5 - 10^7\,M_\odot$).  These ``extreme mass ratio'' binaries
are created when the smaller body is captured onto an extremely strong
field, highly relativistic orbit, generating strong GWs.  Such systems
are measurable to a distance of a few Gigaparsecs if the inspiralling
body is a $10\,M_\odot$ black hole, and to a distance of a few hundred
Megaparsecs if the body is a neutron star or white dwarf.  LISA will
measure the waves that come from the last year or so of the smaller
body's inspiral, and thence probe the nature of the larger black
hole's gravitational field deep within the hole's potential.  The
rates for such events are not too well understood and depend on the
details of stellar dynamics in the cores of galaxies.  Extremely
conservative estimates typically find that the rate of measurable
events for LISA should be at least several per year
{\cite{sigurdssonrees,sigurdsson}}.  Recent thinking suggests that
these rates are likely to be rather underestimated --- black holes
(which are measurable to much greater distances) are likely to
dominate the measured rate, perhaps increasing the rate to several
dozen or several hundred per year {\cite{gair}}.

Finally, it is worth noting that many events involving {\it
intermediate} mass black holes --- those with masses in the band
running from a few $10^2$ to a few $10^5\,M_\odot$ --- would generate
GWs in LISA's sensitive band.  There is a large body of tentative
evidence for the existence of black holes in this mass band (see,
e.g., {\cite{cm05}} for a review), though as yet we have no ``smoking
gun'' unambiguous signature for such a hole.  If such black holes do
exist and undergo mergers in sufficient numbers, measurement of their
waves will make possible a wealth of interesting tests of relativity
{\cite{m05}}, and could untangle some of the mysteries surrounding
supermassive black hole formation and growth.

\subsubsection{Stochastic backgrounds}

As discussed in Sec.\ \ref{sec:stochastic}, ground-based detectors can
measure a stochastic background by correlating the data streams of
widely separated detectors.  LISA will use a slightly different
technique: by combining its six data streams in an appropriate way,
one can construct an observable that is completely {\it in}sensitive
to GWs, measuring noise only {\cite{aet99}}.  This makes it possible
to distinguish between a noise-like stochastic background and true
instrumental noise, and thereby to learn about the characteristics of
the background {\cite{hb2001}}.

The sensitivity of LISA will not be good enough to set interesting
limits on an inflationary GW background: LISA will only reach
$\Omega_{\rm gw} \sim 10^{-11}$, about four orders of magnitude too
large to begin to say something about inflation {\cite{turner1}}.
However, LISA's band is well placed for other possible sources of
cosmological backgrounds.  In particular, an electroweak phase
transition at temperature $T \sim 100 - 1000$ GeV would generate waves
in LISA's band [cf.\ Eq.\ (\ref{eq:freq_phasetrans})].  These waves
are likely to be detectable if the phase transition is strongly first
order (a scenario that does not occur in the standard model, but is
conceivable in extensions to the standard model {\cite{kmk2001}}).

\subsection{Very low frequency}

The very low frequency band, $10^{-9}\,{\rm Hz}\lesssim f\lesssim
10^{-7}\,{\rm Hz}$, corresponds to waves with periods ranging from a
few months to a few decades.  Our best limits on waves in this band
come from observations of millisecond pulsars.  First suggested by
Sazhin {\cite{sazhin}} and then carefully analyzed and formulated by
Detweiler {\cite{det79}}, GWs can drive oscillations in the arrival
times of pulses from a distant pulsar.  Millisecond pulsars are very
good ``detectors'' for measurements in this band because they are
exquisitely precise clocks.  The range of frequencies encompassed by
the very low frequency band is set by the properties of these radio
pulsar measurements: the high end of the frequency band comes from the
need to integrate the radio pulsar data for at least several months;
the low end comes from the fact that we have only been observing
millisecond pulsars for a few decades.  (One cannot observe a
periodicity shorter than the span of one's dataset.)  A recent upper
limit derived from this technique is \cite{mchugh}
\begin{equation}
\Omega_{\rm gw} h_{100}^2 < 9.3\times 10^{-8},\ \ \ \ \ 4\times
10^{-9}~{\rm Hz}< f < 4 \times 10^{-8}~{\rm Hz}
\label{eq:pulsarlimit}
\end{equation}
(where the limit is a $95\%$ confidence limit and $h_{100}$ is the
Hubble constant in units of $100\,{\rm km}\,{\rm sec}^{-1}\,{\rm
Mpc}^{-1}$).

The upper limit (\ref{eq:pulsarlimit}) already places constraints on
some cosmological models (in particular those involving cosmic
strings).  With further observations and the inclusion of additional
pulsars in the datasets, it is likely to improve quite soon.  It is
possible that the background in this band will be dominated by many
unresolved coalescing massive binary black holes {\cite{jb2002}} ---
binaries that are either too massive to radiate in the LISA band, or
else are inspiralling towards the LISA band en route to a final merger
several centuries or millenia hence.  Constraints from pulsar
observations in this band will remain an extremely important source of
data on stochastic waves in the future --- the limits they can set on
$\Omega_{\rm gw}$ are likely to be better than can be set by {\it any}
of the laser interferometric detectors.

\subsection{Ultra low frequency}
\label{subsec:ultralow}

The ultra low frequency band, $10^{-18}\,{\rm Hz}\lesssim f\lesssim
10^{-13}\,{\rm Hz}$, is better described by converting from frequency
to wavelength: For these waves, $10^{-5}\,H_0^{-1}\lesssim \lambda
\lesssim H_0^{-1}$, where $H_0^{-1}\sim 10^{10}$ light years is the
Hubble length today.  Waves in this band oscillate on scales
comparable to the size of the Universe.  They are most likely to be
generated during inflation: Quantum fluctuations in the spacetime
metric are parametrically amplified during inflation to relatively
high amplitude.  The rms amplitude to which the waves are amplified
depends upon the energy scale $E_{\rm infl}$ of inflation:
\begin{equation}
h_{\rm rms} \propto \left(E_{\rm infl}\over m_{\rm P}\right)^2\;,
\end{equation}
where $m_{\rm P}$ is the Planck mass.
Measuring these GWs would be a direct probe of inflationary physics,
and would determine the inflation energy scale, which is currently
unknown to within many orders of magnitude.
These waves have been described as the ``smoking gun''
signature of inflation {\cite{turner2}}.

During inflation, quantum fluctuations impact both the scalar field
which drives inflation (the inflaton $\phi$) and the metric of
spacetime.  There exist independent scalar fluctuations (coupled
fluctuations in the inflaton and scalar-type fluctuations in the
metric) and tensor fluctuations (tensor-type fluctuations in the
metric).  The Fourier modes of these scalar and tensor perturbations
are describable as harmonic oscillators in the expanding Universe
{\cite{kamkos99}}.  Each mode undergoes zero-point oscillations in the
harmonic potential.  However, the potential itself is evolving due to
the expansion of the Universe.  The evolution of this potential
parametrically amplifies these zero-point oscillations, creating
quanta of the field {\cite{bruce96}}.  During inflation, the
Universe's scale factor $a(t)$ grows faster than the Hubble length
$H^{-1}$, and so each mode's wavelength likewise grows faster than the
Hubble length.  The mode's wavelength eventually becomes larger than
the Hubble length, or the mode ``leaves the horizon''.  After
inflation ends the mode subsequently renters the horizon.  For
gravitational perturbations the number of quanta generated in the mode
is proportional to the factor by which the Universe expands between
the two different horizon crossings.  Fluctuations in the inflaton
seed density fluctuations, $\delta\rho(\vec r) = \delta\phi(\vec
r)(\partial V/\partial\phi)$ [where $V(\phi)$ is the potential that
drives the inflaton field].  The tensor-type fluctuations in the
spacetime metric are GWs.

Density fluctuations and GWs both leave an imprint upon the cosmic
microwave background (CMB).  First, each contributes to the CMB
temperature anisotropy.  However, even a perfectly measured map of
temperature anisotropy cannot really determine the contribution of GWs
very well because of {\it cosmic variance}: Since we only have one
Universe to observe, we are sharply limited in the number of
statistically independent influences upon the CMB that we can measure.
Large angular scales are obviously most strongly affected by this
variance, and these scales are the ones on which GW most importantly
impact the CMB {\cite{kamkos98}}.

Fortunately, the scalar and tensor contributions also impact the {\it
polarization} of the CMB.  These two contributions can be detangled
from one another in a model-independent fashion.  This detangling uses
the fact that the polarization tensor $P_{ab}({{\bf \hat n}})$ on the
celestial sphere can be decomposed into tensor harmonics.  These
harmonics come in two flavors, distinguished by their parity
properties: The ``E-modes'' or ``gradient-type'' harmonics
$Y_{(lm)ab}^E({\bf\hat n})$ [which pick up a factor $(-1)^l$ under
${\bf\hat n}\to -{\bf\hat n}$], and the ``B-modes'' or ``curl-type''
harmonics $Y_{(lm)ab}^C({\bf\hat n})$ [which pick up a factor
$(-1)^{l+1}$ under ${\bf\hat n}\to -{\bf\hat n}$].  These harmonics
are constructed by taking covariant derivatives on the sphere of the
``ordinary'' spherical harmonics $Y_{lm}({\bf\hat n})$; see
{\cite{kks97}} for details.  Because scalar perturbations have no
handedness, they {\it only} induce gradient-type polarization.  GWs
induce both gradient- and curl-type polarization.  Thus, an
unambiguous detection of the curl-type polarization would confirm
production of GWs by inflation.  (A caveat is that is that
gravitational lensing can convert E-modes to B-modes; this so-called
``cosmic shear'' ultimately limits the sensitivity to GWs of CMB
polarization studies {\cite{kck02}}.)

\section{Conclusion}

This article has summarized many of the most important topics in the
theory of GWs.  Due to space and time limitations, we sadly were not
able to cover all topics with which students of this field should be
familiar.  In particular, we had hoped to include a discussion of
strong field relativity and GW emission.  We confine ourselves, in
this conclusion, to a (very) brief discussion of important aspects of
this subject for GW science, as well as pointers to the relevant
literature.

Linearized theory as described in Secs.\ {\ref{sec:basic_lin}} and
{\ref{sec:lin_in_curved}} is entirely adequate to describe the
propagation of GWs through our Universe and to model the interaction
of GWs with our detectors.  In some cases, it is even adequate to
describe the emission of waves from a source, as described in Sec.\
{\ref{sec:lin_with_source}} (although for sources with
non-negligible self gravity such as binary star systems
one has to augment linearized theory as described in Sec.\
\ref{sec:quad1}).  However,
many sources have very strong self gravity where the linearized
treatment is completely inadequate.  A variety of formalisms have been
developed to handle these cases.

\begin{itemize}

\item {\it Post-Newtonian (PN) theory.}  PN theory is one of the most
important of these formalisms, particularly for modeling binary
systems.  Roughly speaking, PN theory analyzes sources using an
iterated expansion in two variables: The ``gravitational potential'',
$\phi \sim M/r$, where $M$ is a mass scale and $r$ characterizes the
distance from the source; and velocities of internal motion, $v$.  (In
linearized theory, we assume $\phi$ is small but place no constraints
on $v$.)  Newtonian gravity emerges as the first term in the
expansion, and higher order corrections are found as the expansion is
iterated to ever higher order.  Our derivation of the quadrupole
formula in Sec.\ \ref{sec:quad1} gives the leading order term in the
PN expansion of the emitted radiation.  See Luc Blanchet's recent
review {\cite{lb02}} and references therein for a comprehensive
introduction to and explication of this subject.

\item {\it Numerical relativity.}  Numerical relativity seeks to
directly integrate Einstein's equations on a computer.  Ideally, we
would like to to use a well-understood model of a GW source (e.g., a
binary system in which the field strengths are small enough that it is
well described by post-Newtonian theory) as ``initial data'', and then
numerically evolve the Einstein equations from that point to some
final equilibrium configuration.  The form in which we normally
encounter Einstein's equation in textbooks is not well suited to this
task --- the coordinate freedom of general relativity means that there
is no notion of ``time'' built into the equation $G_{ab} = 8\pi
T_{ab}$.  One must introduce some notion of time for the concept of
``initial data'' to have any meaning.  The 4 dimensions of spacetime
are then split into 3+1 dimensions of space and time.  Having made
this choice, Einstein's equations take on a particular form which is
amenable to numerical computation.

A detailed discussion of numerical relativity is given in the
contribution by Choptuik to this volume {\cite{mchere}}; we also
recommend the reviews by Lehner {\cite{l01}} and by Baumgarte and
Shapiro {\cite{bs03}}.  For the purpose of our present discussion, it
suffices to remark that it has proven to be {\it extremely} difficult
to model some of the most interesting and important GW sources.  In
particular, the final stage of binary black hole mergers --- regarded
by many as the ``Holy Grail'' of numerical relativity --- has proven
to be quite a challenge.

\item {\it Perturbation theory.}  In some cases, GW sources can be
modeled as nearly, but not quite, identical to some exact solution of
the Einstein field equations.  For example, the end state of binary
black hole coalescence must be a single black hole.  As we approach
this final state, the system will be well-modeled as the Kerr black
hole solution, plus some distortion that radiates away.  Another
example is a binary consisting of a stellar mass compact body orbiting
a massive black hole.  The binary's spacetime will be well-described
as a single black hole plus a perturbation due to the captured body.
These cases can be nicely described using {\it perturbation theory}:
We treat the spacetime as some exact background, $g_{ab}^{\rm B}$,
plus a perturbation $h_{ab}$:
\begin{equation}
g_{ab} = g^{\rm B}_{ab} + h_{ab}\;.
\label{eq:metricpert}
\end{equation}
We are in the perturbative regime if $||h_{ab}||/||g^{\rm B}_{ab}||
\ll 1$.  This system can then be analyzed by expanding the Einstein
equations for this metric and keeping terms to first order in $h_{ab}$
(see Sec.\ \ref{sec:pertcurved} for details but without the matter
source terms included).

This approach has proven to be particularly fruitful when the
background spacetime is that of a black hole.  For the case of a
Schwarzschild background, the derivation of the full perturbation
equations is rather straightforward; Rezzolla gives a particularly
compact and readable summary {\cite{lr03}}.  Perturbations of Kerr
black holes are not nearly so simple to describe, largely due to the
lack of spherical symmetry --- expanding the metric as in Eq.\
(\ref{eq:metricpert}) does not prove to be so fruitful as it is in the
Schwarzschild case.  Somewhat miraculously, it turns out that progress
can be made by expanding the {\it curvature} tensor: By expanding the
Riemann tensor as $R_{abcd} = R^{\rm B}_{abcd} + \delta R_{abcd}$ and
taking an additional derivative of the Bianchi identity,
\begin{equation}
\nabla_e R_{abcd} + \nabla_d R_{abec} + \nabla_c R_{abde} = 0\;,
\end{equation}
one can derive a wave-like equation for the perturbation $\delta
R_{abcd}$.  This analysis was originally performed by Teukolsky; see
his original analysis {\cite{t73}} for details.

\end{itemize}

\ack

We thank Richard Price and Jorge Pullin for the invitation to write
this article, and are profoundly grateful to Tim Smith at the
Institute of Physics for patiently and repeatedly extending our
deadline as we wrote this article.  We are grateful to two anonymous
referees for detailed and helpful comments, as well as to helpful
comments from Hongbao Zhang.  SAH thanks the Caltech TAPIR group and
the Kavli Institute for Theoretical Cosmology at the University of
Chicago for their hospitality while this article was completed.  EEF
is supported by NSF Grant PHY-0140209; SAH is supported by NSF Grant
PHY-0244424 and NASA Grant NAGW-12906;

\appendix

\section{Existence of TT gauge in local vacuum regions in linearized
  gravity}
\label{sec:localTT}

In this appendix we show that one can always find TT gauges in local
vacuum regions in linearized gravity.  More precisely, suppose that
${\cal V}$ is a connected open spatial region, and $(t_0,t_1)$ is an
open interval of time.  Then one can find a gauge on the product
${\cal R} \equiv (t_0,t_1) \times {\cal V}$ that satisfies $h_{tt} =
h_{ti} = \delta^{ij} h_{ij} = \partial_i h_{ij} =0$, as long as
$T_{ab} =0$ throughout ${\cal R}$.

The proof involves a generalization of the gauge-invariant formalism
of Sec.\ \ref{subsec:gauge_invar} to finite spacetime regions.  We
define a decomposition of the metric perturbation $h_{ab}$ in terms of
quantities $\phi$, $\beta_i$, $\gamma$, $h_{ij}^{\rm TT}$, $H$,
$\varepsilon_i$ and $\lambda$ using the same equations
(\ref{eq:htt_invar}) -- (\ref{eq:constraint4}) as before.  However we
replace the boundary conditions (\ref{eq:boundary}) with
\begin{eqnarray}
\label{eq:boundary2a}
\gamma_{|\partial {\cal V}} &=& \int_{t_0}^t dt \, \phi_{|\partial V}, \\
\lambda_{|\partial {\cal V}} &=& 0, \\
(\nabla^2 \lambda)_{|\partial {\cal V}} &=& H_{|\partial {\cal V}}, \\
({\bf n} \times {\mbox{\boldmath $\varepsilon$}})_{|\partial {\cal V}}
&=&
2 \int_{t_0}^t dt \, ({\bf n} \times {\mbox{\boldmath
$\beta$}})_{|\partial {\cal V}},
\label{eq:boundary2}
\end{eqnarray}
where ${\bf n}$ is the unit outward-pointing unit normal to $\partial
{\cal V}$.  The reason for this particular choice of boundary
conditions will be explained below.  These boundary conditions define
a unique decomposition of the metric within ${\cal R}$.

Next, we compute how the variables $\phi$, $\beta_i$, $\gamma$,
$h_{ij}^{\rm TT}$, $H$, $\varepsilon_i$ and $\lambda$ transform under
general gauge transformations.  We use the same parameterization
(\ref{eq:gauge_functions}) of the gauge transformation as before,
except we impose now the boundary condition $C_{|\partial {\cal V}}
=0$.  We find that the transformation laws (\ref{eq:phi_gauge}) --
(\ref{eq:hTT_gauge}) are replaced by the following equations which
contain some extra terms:
\begin{eqnarray}
\phi &\to& \quad \phi - \dot A\;,
\label{eq:phi_gauge1}\\
\beta_i &\to& \quad \beta_i - \dot B_i - \partial_i \psi \;,
\label{eq:beta_gauge1}\\
\gamma &\to& \quad \gamma - A - \dot C + \psi \;,
\label{eq:gamma_gauge1}\\
H &\to& \quad H - 2\nabla^2C\;,
\label{eq:h_gauge1}\\
\lambda &\to& \quad \lambda - 2 C\;,
\label{eq:lambda_gauge1}\\
\varepsilon_i &\to& \quad \varepsilon_i - 2 B_i + 2 \eta_i - 2 (t-t_0)
\partial_i \psi \;,
\label{eq:epsilon_gauge1}\\
h_{ij}^{\rm TT} &\to& \quad h_{ij}^{\rm TT} - 2 \partial_{(i}
\eta_{j)} + 2 (t - t_0) \partial_i \partial_j \psi \;.
\label{eq:hTT_gauge1}
\end{eqnarray}
Here $\psi$ is the time-independent, harmonic function defined by
$\nabla^2 \psi =0$ and $\psi_{|\partial {\cal V}} = A_{|\partial {\cal
V},t=t_0}$.  Similarly $\eta_i$ is the time-independent, harmonic
transverse vector defined by $\nabla^2 \eta_i=0$ and
$( {\bf n} \times {\mbox{\boldmath $\eta$}})_{|\partial {\cal V}} =
  {\bf n} \times {\bf B}_{|\partial {\cal
 V},t=t_0}.
%\end{equation}
$

We define the variables $\Phi$, $\Theta$ and $\Xi_i$ by the same
equations (\ref{eq:Phidef}) -- (\ref{eq:Xidef}) as before.  From the
transformation laws (\ref{eq:phi_gauge1}) -- (\ref{eq:hTT_gauge1})
these variables are still gauge invariant, while $h_{ij}^{\rm TT}$ is
no longer gauge invariant in the present context.  Next, imposing the
linearized vacuum Einstein equations using the expressions
(\ref{eq:Gtt}) -- (\ref{eq:Gij}) yields
\begin{equation}
\nabla^2 \Theta =0, \ \ \ \
\nabla^2 \Xi_i = -2 \partial_i {\dot \Theta}, \ \ \ \
\nabla^2 \Phi = \frac{3}{2} {\ddot \Theta}
\end{equation}
in ${\cal V}$.  The boundary conditions (\ref{eq:boundary2a}) --
(\ref{eq:boundary2}) together with the definitions (\ref{eq:Phidef})
-- (\ref{eq:Xidef}) imply that the boundary conditions on the gauge
invariant variables are
\begin{equation}
\Phi_{|\partial {\cal V}} = \Theta_{|\partial {\cal V}} =
\Xi^i_{|\partial {\cal V}} =0.
\label{eq:vanish}
\end{equation}
(This is why we choose those particular boundary conditions.)
Therefore all the gauge invariant variables vanish, $\Theta = \Phi =
\Xi^i =0$ in ${\cal R}$.

It is now straightforward to show that one can choose a gauge in which
$\phi = \beta_i  = \gamma = H = \varepsilon_i = \lambda = 0$.
From the transformation laws
(\ref{eq:phi_gauge1}) -- (\ref{eq:hTT_gauge1})
we can choose $C$ to make $\lambda=0$, choose ${\dot A}$ to make $\phi
=0$, and choose ${\dot B}_i$ to make $\beta_i=0$.
The residual gauge freedom is then parameterized by functions $A$ and
$B_i$ that are time-independent.  Next, from Eq.\ (\ref{eq:vanish})
together with the definitions (\ref{eq:Phidef}) -- (\ref{eq:Xidef}) it
follows that
\begin{eqnarray}
0 &=& \Theta = \frac{1}{3} H \\
0 &=& \Phi = - 2 {\dot \gamma} \\
0 &=& \Xi_i = - \frac{1}{2} {\dot \varepsilon}_i.
\end{eqnarray}
Thus the only remaining non-zero pieces of the metric other than the
TT piece are $\gamma$ and $\varepsilon_i$, and these are both
time-independent.  Finally we can use the residual gauge freedom given
by time-independent functions $A$ and $B_i$ to set to zero $\gamma$
and $\varepsilon_i$, by Eqs.\ (\ref{eq:gamma_gauge1}) and
(\ref{eq:epsilon_gauge1}).  [For this purpose $A$ and $B_i$ will
  vanish on $\partial {\cal V}$, by
Eqs.\ (\ref{eq:boundary2a}) and (\ref{eq:boundary2}), so $\psi$ and
$\eta_i$ vanish.]

\end{document}